\documentclass[journal]{IEEEtran}
\usepackage{cite}
\usepackage{amsmath}
\usepackage{amssymb}
\usepackage{amsfonts}
\usepackage{graphicx}
\usepackage{graphics}
\usepackage{textcomp}

\usepackage{scalerel}
\usepackage{tikz}
\usepackage{tikz-3dplot}
\usepackage[utf8]{inputenc}
\usepackage{pgfplots}
\pgfplotsset{compat=newest}
\usepgfplotslibrary{groupplots}
\usepgfplotslibrary{dateplot}
\usetikzlibrary{svg.path}
\definecolor{orcidlogocol}{HTML}{A6CE39}
\tikzset{
orcidlogo/.pic={
\fill[orcidlogocol] svg{M256,128c0,70.7-57.3,128-128,128C57.3,256,0,198.7,0,128C0,57.3,57.3,0,128,0C198.7,0,256,57.3,256,128z};
\fill[white] svg{M86.3,186.2H70.9V79.1h15.4v48.4V186.2z}
svg{M108.9,79.1h41.6c39.6,0,57,28.3,57,53.6c0,27.5-21.5,53.6-56.8,53.6h-41.8V79.1z M124.3,172.4h24.5c34.9,0,42.9-26.5,42.9-39.7c0-21.5-13.7-39.7-43.7-39.7h-23.7V172.4z}
svg{M88.7,56.8c0,5.5-4.5,10.1-10.1,10.1c-5.6,0-10.1-4.6-10.1-10.1c0-5.6,4.5-10.1,10.1-10.1C84.2,46.7,88.7,51.3,88.7,56.8z};
}
}
\newcommand\orcidicon[1]{\href{https://orcid.org/#1}{\mbox{\scalerel*{
\begin{tikzpicture}[yscale=-1,transform shape]
\pic{orcidlogo};
\end{tikzpicture}
}{|}}}}

\usepackage{hyperref}

\usepackage{amsmath}
\usepackage{alphalph}
\usepackage{etoolbox}
\makeatletter
\newalphalph{\aalphalph}[mult]{\alphalph@alph}{26}
\newcommand{\alphalphval}[1]{%
  \@ifundefined{c@#1}{
    \aalphalph{#1}
  }{%
    \aalphalph{\value{#1}}
  }
}
\makeatother

\AtBeginEnvironment{subequations}{%
  \let\alph\alphalphval%
}

\usepackage{xcolor}
\usepackage[utf8]{inputenc}
\usepackage[english]{babel}
\usepackage{amsmath}
\usepackage{amssymb}
\usepackage{array}
\usepackage{eqparbox}
\usepackage{url}
\usepackage{amsfonts}
\usepackage{mathtools}
\usepackage{multicol}
\usepackage{booktabs}

\usepackage{textcomp}
\usepackage{pifont}
\newcommand{\xmark}{\text{\ding{55}}}

\DeclareUnicodeCharacter{2009}{-}

\newcommand{\p}{\textup{\texttt{+}}} % small unary plus
\newcommand{\m}{\textup{\texttt{-}}} % small unary minus
\usepackage{comment}
\usepackage{array}
\usepackage{eqparbox}
\usepackage{url}
\usepackage{amsfonts}
\usepackage{alphalph,etoolbox}
\allowdisplaybreaks
\patchcmd{\subequations}{\alph{equation}}{\alphalph{\value{equation}}}{}{}
\usepackage{multirow}
\usepackage{graphics}
\usepackage{textcomp}
\usepackage{amsmath}
\usepackage{comment}
\usepackage{amsfonts}
\usepackage{amsthm}
\usepackage{cite}
\usepackage{algpseudocode}
\usepackage{algorithm}
\usepackage{url}
\usepackage{etoolbox}
\usepackage[official]{eurosym}
\usepackage{cancel}
\definecolor{Gray}{gray}{0.9}
\definecolor{LightCyan}{rgb}{0.88,1,1}
\usepackage{tikz}
\usepackage{pgfplots}
\usetikzlibrary{shapes,arrows}
\usepackage{tikz-3dplot}
\pgfplotsset{compat=newest}
\usepgfplotslibrary{groupplots}
\usepgfplotslibrary{dateplot}
\usepackage{svg}
\allowdisplaybreaks%Solve problem of equation breaking between text
\usepackage{colortbl}%rowcolor
\usepackage{tabularx}
\newcolumntype{Y}{>{\centering\arraybackslash}X}
\usepackage{url}
\usepackage[flushleft]{threeparttable}
\pgfplotsset{width=7cm,compat=1.12}
\usepgfplotslibrary{statistics}

\definecolor{color0}{rgb}{0.917647058823529,0.917647058823529,0.949019607843137}
\definecolor{color1}{rgb}{0.298039215686275,0.447058823529412,0.690196078431373}
\definecolor{color2}{rgb}{0.333333333333333,0.658823529411765,0.407843137254902}
\definecolor{color3}{rgb}{0.768627450980392,0.305882352941176,0.32156862745098}
\definecolor{color4}{rgb}{0.505882352941176,0.447058823529412,0.698039215686274}

\definecolor{color10}{rgb}{0.917647058823529,0.917647058823529,0.949019607843137}
\definecolor{color11}{rgb}{0.748039215686275,0.700980392156863,0.812745098039216}
\definecolor{color12}{rgb}{0.933823529411765,0.754411764705882,0.583823529411765}
\definecolor{color13}{rgb}{0.95,0.95,0.65}
\definecolor{color14}{rgb}{0.27843137254902,0.431372549019608,0.631372549019608}
\definecolor{color15}{rgb}{0.824509803921569,0.124509803921569,0.492156862745098}
\definecolor{color16}{rgb}{0.666666666666667,0.372549019607843,0.172549019607843}

\definecolor{color100}{rgb}{0.917647058823529,0.917647058823529,0.949019607843137}
\definecolor{color101}{rgb}{0.298039215686275,0.447058823529412,0.690196078431373}
\definecolor{color102}{rgb}{0.333333333333333,0.658823529411765,0.407843137254902}
\definecolor{color103}{rgb}{0.768627450980392,0.305882352941176,0.32156862745098}
\definecolor{color104}{rgb}{0.505882352941176,0.447058823529412,0.698039215686274}
\definecolor{color110}{rgb}{0.917647058823529,0.917647058823529,0.949019607843137}
\definecolor{color111}{rgb}{0.748039215686275,0.700980392156863,0.812745098039216}
\definecolor{color112}{rgb}{0.933823529411765,0.754411764705882,0.583823529411765}
\definecolor{color113}{rgb}{0.95,0.95,0.65}
\definecolor{color114}{rgb}{0.27843137254902,0.431372549019608,0.631372549019608}
\definecolor{color115}{rgb}{0.824509803921569,0.124509803921569,0.492156862745098}
\definecolor{color116}{rgb}{0.666666666666667,0.372549019607843,0.172549019607843}

\definecolor{darkgray176}{RGB}{176,176,176}
\definecolor{darkorange25512714}{RGB}{255,127,14}
\definecolor{forestgreen4416044}{RGB}{44,160,44}
\definecolor{gray}{RGB}{128,128,128}
\definecolor{lightgray204}{RGB}{204,204,204}
\definecolor{steelblue31119180}{RGB}{31,119,180}

\usepackage{nomencl}
\makenomenclature
\renewcommand\nomgroup[1]{%
  \item[\bfseries
  \ifstrequal{#1}{I}{Indices}{%
      \ifstrequal{#1}{V}{Variables (lower-case letters)}{%
          \ifstrequal{#1}{C}{Constants}{%
              \ifstrequal{#1}{P}{Parameters (upper-case letters)}{%
                \ifstrequal{#1}{J}{Sets}{}%
              }%
          }%
      }%
  }%
]}
\usepackage{listings}
\usepackage[numbered,framed]{matlab-prettifier}
\lstdefinelanguage{matlab}{
  style              = Matlab-editor,
  basicstyle         = \mlttfamily,
  escapechar         = ",
  mlshowsectionrules = true,
}
\lstdefinelanguage{GAMS}{
morekeywords={
ABORT , ACRONYM , ACRONYMS , ALIAS , ALL , AND , ASSIGN , BINARY , CARD , DISPLAY , EPS , EQ , EQUATION , EQUATIONS , GE , GT , INF , INTEGER , LE , LOOP , LT , MAXIMIZING , MINIMIZING , MODEL , MODELS , NA , NE , NEGATIVE , NOT , OPTION , OPTIONS , OR , ORD , PARAMETER , PARAMETERS , POSITIVE , PROD , SCALAR , SCALARS , SET , SETS , SMAX , SMIN , SOS1 , SOS2 , SUM , SYSTEM , TABLE , USING , VARIABLE , VARIABLES , XOR , YES , REPEAT , UNTIL , WHILE , IF , THEN , ELSE , SEMICONT , SEMIINT , FILE , FILES , PUT , PUTPAGE , PUTTL , PUTCLOSE , FREE , NO , SOLVE , FOR , ELSEIF , ABS , ARCTAN , CEIL , COS , ERROR , EXP , FLOOR , LOG , LOG10 , MAP , MAPVAL , MAX , MIN , MOD , NORMAL , POWER , ROUND , SIGN , SIN , SQR , SQRT , TRUNC , UNIFORM , LO , UP , FX , SCALE , PRIOR , PC , PS , PW , TM , BM , CASE , DATE , IFILE , OFILE , PAGE , RDATE , RFILE , RTIME , SFILE , TIME , TITLE , TS , TL , TE , TF , LJ , NJ , SJ , TJ , LW , NW , SW , TW , ND , NR , NZ , CC , HDCC , TLCC , LL , HDLL , TLLL , LP , WS , /,PROD: },
sensitive = false,
morecomment=[f]*,%
morecomment=[s]{$ontext$}{$offtext$},
morestring=[b]”,
morestring=[b]’
}
\usetikzlibrary{shapes.geometric}

\newcommand{\labDA}{\mathrm{D\!A\!}}
\newcommand{\labFC}{\mathrm{F\!C\!}}
\newcommand{\labID}{\mathrm{I\!D\!}}
\newcommand{\labF}{\mathrm{F}}
\newcommand{\labG}{\mathrm{G}}
\newcommand{\labST}{\mathrm{S\!T}}
\newcommand{\labTH}{\mathrm{T\! H}}
\newcommand{\conNBS}{\mathrm{N\!B}}
\newcommand{\conNHS}{\mathrm{N\!H}}
\newcommand{\conNL}{\mathrm{N\!L}}
\newcommand{\conNN}{\mathrm{N\!N}}
\newcommand{\conNNST}{\mathrm{N\!N}^{(\!\m \labST\!)}}
\newcommand{\conNNSTTH}{\mathrm{N\!N}^{(\!\labTH\!)}}
\newcommand{\conNST}{\mathrm{N\!N}^{(\!\labST\!)}}
\newcommand{\conNT}{\mathrm{N\!T}}
\newcommand{\conNW}{\mathrm{N\!W}}
\newcommand{\indj}{j}%node
\newcommand{\indk}{k}%hydro segment
\newcommand{\indl}{l}%line
\newcommand{\indn}{n}%node/area
\newcommand{\inds}{s}%bid segment
\newcommand{\indw}{w}%scenario
\newcommand{\indt}{t}%time

\newcommand{\setK}{\mathbb{K}}
\newcommand{\setL}{\mathbb{L}}
\newcommand{\setS}{\mathbb{S}}
\newcommand{\setN}{\mathbb{N}}
\newcommand{\setNNST}{\mathbb{N}^{\!\,\m \!\labST\!}}
\newcommand{\setNST}{\mathbb{N}^{\,\!\labST\!}}%strategic
\newcommand{\setNNSTTH}{\mathbb{N}^{\,\labTH\!}}
\newcommand{\setW}{\mathbb{W}}
\newcommand{\setT}{\mathbb{T}}
\newcommand{\setXup}{\mathbb{X}^{U\!P\!}}
\newcommand{\setXDA}{\mathbb{X}^{\labDA}}
\newcommand{\setXFC}{\mathbb{X}^{\labFC}}

\newcommand{\setYDA}{\mathbb{Y}^{\labDA}_{\indw}}

\newcommand{\setYFC}{\mathbb{Y}^{\labFC}_{\indw}}

\newcommand{\parAU}{A^{\!U}_{\indn\indj}}
\newcommand{\parAD}{A^{\!D}_{\indn\indj}}
\nomenclature[Paq]{$\parAU/\parAD$}{Adjacency matrix for hydrologically connected nodes($\indj$) upstream/downstream of node $\indn$;}
\newcommand{\parAL}{A^{\!L}_{\indl\indn}}

\nomenclature[PAL]{$\parAL$}{Adjacency matrix value; 
}
\newcommand{\parBIDhDAmax}{\overline{B}^{\labDA}_{\indn\inds\indt}}

\nomenclature[PbhDAmax]{$\parBIDhDAmax$}{Maximum bid price
in DA market (\EUR{}/MWh);}
\newcommand{\parBIDhDAmin}{\underline{B}^{\labDA}_{\indn\inds\indt}}

\nomenclature[PbhDAmin]{$\parBIDhDAmin$}{Minimum bid price
in DA market (\EUR{}/MWh);}
\newcommand{\parBIDhFCmax}{\overline{B}^{\labFC}_{\indn\inds\indt}}

\nomenclature[PbhFCmax]{$\parBIDhFCmax$}{Maximum bid price
in FCR-N market (\EUR{}/MW);}
\newcommand{\parBIDhFCmin}{\underline{B}^{\labFC}_{\indn\inds\indt}}

\nomenclature[PbhFCmin]{$\parBIDhFCmin$}{Minimum bid price
in FCR-N market (\EUR{}/MW);}
\newcommand{\parVH}{V_{\indn\indt\indw}}
\newcommand{\parVHindnone}{V_{1\indt}}
\newcommand{\parVHindntwo}{V_{2\indt}}

\nomenclature[PVH]{$\parVH$}{Local inflow 
($m^3$);}
\newcommand{\parGHmax}{\overline{P}^{}_{\indn}}

\newcommand{\IparGHmaxindnthree}{\overline{P}^{}_{3}}

\newcommand{\parGHmin}{\underline{P}^{}_{\indn}}
\nomenclature[PGHmax]{$\parGHmax$/$\parGHmin$}{Maximum/minimum generation capacity 
(MW);}
\newcommand{\parMHmax}{\overline{M}^{}_{\indn}}

\newcommand{\parMHmin}{\underline{M}^{}_{\indn}}

\nomenclature[PMHmax]{$\parMHmax$/$\parMHmin$}{Maximum/minimum reservoir
($m^3$);}
\newcommand{\parSHmax}{\overline{S}^{}_{\indn}}
\newcommand{\parSHmin}{\underline{S}^{}_{\indn}}
\nomenclature[PSHmax]{$\parSHmax$/$\parSHmin$}{Maximum/minimum spillage
($m^3$);}
\newcommand{\parQHmax}{\overline{Q}^{}_{\indk\indn}}

\newcommand{\parQHmin}{\underline{Q}^{}_{\indk\indn}}
\nomenclature[PQHmax]{$\parQHmax$/$\parQHmin$}{Maximum/minimum discharge
($m^3$);}
\newcommand{\parGHIDmax}{\overline{P}^{\!I\!D\p\!}_{\indn}}

\nomenclature[PGHIDmax]{$\parGHIDmax$}{Maximum participation level in intraday (MWh);}

\newcommand{\parGHFChmin}{\underline{P}^{\labFC}_{\indn}}
\nomenclature[PGHFChmin]{$\parGHFChmin$}{Minimum participation level
in FCR-N market (MW);}
\newcommand{\parglNTCmax}{\overline{C}_{\indl}}
\newcommand{\IparglNTCmaxindlone}{\overline{C}_{1}}
\newcommand{\IparglNTCmaxindltwo}{\overline{C}_{2}}
\nomenclature[PglNTCmax]{$\parglNTCmax$}{Maximum net transfer capacity
$\indl$ (MW);}
\newcommand{\parDDA}{D^{\labDA}_{\indn\indt\indw}}

\nomenclature[PDDA]{$\parDDA$}{Demand in the DA market
(MWh);}
\newcommand{\parDFC}{D^{\labFC}_{\indt\indw}}

\nomenclature[PDFC]{$\parDFC$}{Demand in FCR-N market
(MW);}
\newcommand{\parDID}{D^{\labID}_{\indn\indt\indw}}
\nomenclature[PDID]{$\parDID$}{Demand in ID market (MW);}    
\newcommand{\parlambdaf}{\lambda^{\labF}_{\indw}}
\nomenclature[Plambdaf]{$\parlambdaf$}{The future electricity price (\EUR{}/MWh);}
\newcommand{\parmuh}{\mu_{\indk\indn}}

\nomenclature[Pmuh]{$\parmuh$}{Production equivalent (MWh/$m^3$);}
\newcommand{\partauh}{\tau_{\indj}}
\nomenclature[Ptauh]{$\partauh$}{Water flow time to downstream station ($h$);}
\newcommand{\parctDA}{c^{\labDA}_{\indn}}

\nomenclature[PctDA]{$\parctDA$}{Thermal cost
in DA market (\EUR{}/MWh);}
\newcommand{\parctFC}{c^{\labFC}_{\indn\indt\indw}}

\nomenclature[PctFC]{$\parctFC$}{Thermal cost
in FCR-N market (\EUR{}/MW);}

\newcommand{\parlambdahIDp}{\lambda^{\!I\!D\p\!}_{\indn\indt\indw}}

\newcommand{\parlambdahIDm}{\lambda^{\!I\!D\m\!}_{\indn\indt\indw}}
\nomenclature[PlambdahIDp]{$\parlambdahIDp$/$\parlambdahIDm$}{ID market prices (\EUR{}/MWh);}
\newcommand{\parpammah}{\mu^{\labF}_{\indj}}
\nomenclature[Pgammah]{$\parpammah$}{Expected future production equivalent}
\newcommand{\parprob}{\pi_{\indw}}
\nomenclature[Pprob]{$\parprob$}{Probability of scenario $\indw$;}
\newcommand{\pardelta}{\delta_{\indn}}

\nomenclature[Pdelta]{$\pardelta$}{Maximum value of droop setting (MW/Hz);}
\newcommand{\pardeltaf}{\Delta f_{\indt}}

\nomenclature[Pdeltaf]{$\pardeltaf$}{Frequency deviation (Hz);}

\newcommand{\parmthetaseven}{M^{17}}

\newcommand{\parmthetaten}{M^{20}}
\newcommand{\parpricecapDA}{200}
\newcommand{\parpricecapFC}{100}
\newcommand{\varphDA}{g^{\labDA}_{\indn\indt\indw}}

\nomenclature[VghDA]{$\varphDA$}{Dispatched power by Non-Strategic (NST) unit
in DA market (MWh);}
\newcommand{\varphDAseg}{p^{\labDA}_{\indn\inds\indt\indw}}
\newcommand{\varphDAsegindnone}{p^{\labDA}_{1\indt}}

\nomenclature[VghDAseg]{$\varphDAseg$}{Dispatched power by Strategic (ST) unit
in DA market (MWh);}
\newcommand{\parphG}{p^{\labG}_{\indn\indt\indw}}

\nomenclature[VghG]{$\parphG$}{Generated power by strategic hydro unit $\indn$;}
\newcommand{\varphDAseghat}{\hat{p}^{\labDA}_{\indn\inds\indt\indw}}
\newcommand{\varpBIDhDA}{p^{b\labDA}_{\indn\inds\indt}}
\nomenclature[VgbhDA]{$\varpBIDhDA$}{Quantity offer of hydro unit
in DA market (MW);}
\newcommand{\varpBIDhFC}{p^{b\labFC}_{\indn\inds\indt}}

\nomenclature[VgbhFC]{$\varpBIDhFC$}{Quantity offer of hydro unit
in FCR-N market (MW);}
\newcommand{\parpBIDhDAhat}{\hat{p}^{b\labDA}_{\indn\inds\indt}}

\newcommand{\parpBIDhFChat}{\hat{p}^{b\labFC}_{\indn\inds\indt}}

\newcommand{\varphFC}{g^{\labFC}_{\indn\indt\indw}}

\nomenclature[VghFC]{$\varphFC$}{Dispatched power by NST unit
in FCR-N (MW);}
\newcommand{\varphFCseg}{p^{\labFC}_{\indn\inds\indt\indw}}

\nomenclature[VghFCseg]{$\varphFCseg$}{Dispatched power by ST unit in FCR-N (MW);}
\newcommand{\varphIDp}{p^{\!I\!D\p\!}_{\indn\indt\indw}}

\newcommand{\varphIDm}{p^{\!I\!D\m\!}_{\indn\indt\indw}}

\nomenclature[VghIDp]{$\varphIDp$/$\varphIDm$}{Dispatched power
in ID market (MWh);}
\newcommand{\varphDAhat}{\hat{g}^{\labDA}_{\indn\indt\indw}}

\newcommand{\varplDA}{p^{\!L}_{\indl\indt\indw}}

\nomenclature[VglDA]{$\varplDA$}{Power of line $\indl$ (MW);}    
\newcommand{\varqhDA}{q^{\labDA}_{\indn\indt\indw}}
\nomenclature[VqhDA]{$\varqhDA$}{Hydro reservoir content for DA market ($m^3$);} 
\newcommand{\varBIDhDAseg}{b^{\labDA}_{\indn\inds\indt}}

\nomenclature[VbhDAseg]{$\varBIDhDAseg$}{Price bid of hydro unit in DA market (\EUR{}/MWh);}
\newcommand{\varBIDhDAsegminusone}{b^{\labDA}_{\indn,\inds\m1,\indt}}
\newcommand{\varBIDhFCseg}{b^{\labFC}_{\indn\inds\indt}}

\newcommand{\parBIDhFCseghat}{\hat{b}^{\labFC}_{\indn\inds\indt}}

\nomenclature[VbhFCseg]{$\varBIDhFCseg$}{Price bid of hydro unit
in FCR-N market
(\EUR{}/MW);}
\newcommand{\varBIDhFCsegminusone}{b^{\labFC}_{\indn,\inds\m1,\indt}}
\newcommand{\vardh}{q^{}_{\indk\indn\indt\indw}}

\nomenclature[Vdh]{$\vardh$}{Discharge volume 
($m^3$);}

\newcommand{\vardhminustau}{q^{}_{\indk\indj,\indt\m\partauh,\indw}}

\newcommand{\parBIDhDAseghat}{\hat{b}^{\labDA}_{\indn\inds\indt}}

\newcommand{\varduallambdahDA}{\lambda^{\labDA}_{\indn\indt\indw}}

\newcommand{\IvarduallambdahDA}{\lambda^{\labDA}}
\newcommand{\IvarduallambdahDAindnone}{\lambda^{\labDA}_{1}}
\newcommand{\IvarduallambdahDAindntwo}{\lambda^{\labDA}_{2}}

\newcommand{\varduallambdahFC}{\lambda^{\labFC}_{\indt\indw}}

\newcommand{\IvarduallambdahFC}{\lambda^{\labFC}_{}}
\newcommand{\varmh}{m^{}_{\indn\indt\indw}}

\newcommand{\varmhzero}{m^{}_{\indn0\indw}}
\newcommand{\varmhT}{m^{}_{\indn T \indw}}
\nomenclature[Vmh]{$\varmh$}{Reservoir content ($m^3$);}
\newcommand{\varmhminusone}{m^{}_{\indn,\indt\m1,\indw}}
\newcommand{\varsh}{s^{}_{\indn\indt\indw}}

\nomenclature[Vsh]{$\varsh$}{Spillage ($m^3$);}
\newcommand{\varshminustau}{s^{}_{\indj,\indt\m\partauh,\indw}}
\newcommand{\vardualthetaone}{\theta^{1}_{\indn\inds\indt\indw}}

\newcommand{\vardualthetasix}{\theta^{6}_{\indn\indt\indw}}

\newcommand{\vardualthetaseven}{\theta^{7}_{\indn\indt\indw}}

\newcommand{\vardualthetanine}{\theta^{9}_{\indn\inds\indt\indw}}

\newcommand{\vardualthetaten}{\theta^{10}_{\indn\indt\indw}}

\newcommand{\vardualetaone}{\eta^{1}_{\indn\indt\indw}}

\newcommand{\vardualnutwoover}{\overline{\nu}^{2}_{\indk\indn\indt\indw}}
\newcommand{\vardualnutwounder}{\underline{\nu}^{2}_{\indk\indn\indt\indw}}
\newcommand{\vardualnuthreeover}{\overline{\nu}^{3}_{\indn\indt\indw}}
\newcommand{\vardualnuthreeunder}{\underline{\nu}^{3}_{\indn\indt\indw}}
\newcommand{\vardualnufourover}{\overline{\nu}^{4}_{\indn\indt\indw}}
\newcommand{\vardualnufourunder}{\underline{\nu}^{4}_{\indn\indt\indw}}
\newcommand{\vardualnufiveover}{\overline{\nu}^{5}_{\indl\indt\indw}}

\newcommand{\vardualnufiveunder}{\underline{\nu}^{5}_{\indl\indt\indw}}

\newcommand{\vardualnusixover}{\overline{\nu}^{6}_{\indn\inds\indt\indw}}

\newcommand{\vardualnusixunder}{\underline{\nu}^{6}_{\indn\inds\indt\indw}}

\newcommand{\vardualnusevenover}{\overline{\nu}^{7}_{\indn\indt\indw}}

\newcommand{\vardualnusevenunder}{\underline{\nu}^{7}_{\indn\indt\indw}}

\newcommand{\vardualetatwo}{\eta^{2}_{\indn\indt\indw}}

\newcommand{\expEone}{E^{(1)}_{\indn\inds\indt\indw}}
\newcommand{\expEtwo}{E^{(2)}_{\indn\inds\indt\indw}}
\newcommand{\expEfour}{E^{(3)}_{\indn\indt\indw}}

\newcommand{\expEfive}{E^{(4)}_{\indn\indt\indw}}

\newcommand{\nutpardelta}{n}%{\delta}%desired mean acceptable probability
\nomenclature[Pnutpardelta]{$\nutpardelta$}{No U-Turn sampler (NUTS)-based algorithm: desired mean;}
\newcommand{\nutparlog}{L}%{\mathcal{P}}%Logarithm of the join density
\nomenclature[Pnutparlog]{$\nutparlog$}{NUTS-based algorithm: target distribution function;}
\newcommand{\nutparM}{N\!S}%{N\!P}%number of samples/steps
\nomenclature[PnutparM]{$\nutparM$}{NUTS-based algorithm: number of iterations;}
\newcommand{\nutparMa}{N\!S^{a}}%{N\!P^{a}}%number of samples to adapt
\nomenclature[PnutparMa]{$\nutparMa$}{NUTS-based algorithm: number of iterations to adapt;}
\newcommand{\nutdefN}{N}%{\mathcal{N}}
\nomenclature[PnutparN]{$\nutdefN$}{NUTS-based algorithm: normal distribution function;}
\newcommand{\nutdefuniform}{u}%{\text{uni}}
\nomenclature[Pnutparuni]{$\nutdefuniform$}{NUTS-based algorithm: uniform distribution function;}

\newcommand{\nutdefstep}{S}%{\mathcal{U}}
\nomenclature[PnutU]{$\nutdefstep$}{NUTS-based algorithm: step function;}

\newcommand{\nutvartheta}{x}%{\textbf{x}}
\nomenclature[Vnutvarx]{$\nutvartheta$}{NUTS-based algorithm: position;}
\newcommand{\nutvarr}{\eta}%{\boldsymbol{\rho}}
\nomenclature[Vnutvarr]{$\nutvarr$}{NUTS-based algorithm: momentum;}
\newcommand{\nutvarepsilon}{\tau}%{\xi}
\nomenclature[Vnutvarxi]{$\nutvarepsilon$}{NUTS-based algorithm: step size;}
\begin{document}
\algnewcommand{\algorithmicgoto}{\textbf{go to}}%
\algnewcommand{\Goto}[1]{\algorithmicgoto~\ref{#1}}%

\title{Probabilistic Multi-product Trading in Sequential Intraday and Frequency-Regulation Markets}
\author{Saeed~Nordin\orcidicon{0000-0003-1823-9653},~\IEEEmembership{Student Member,~IEEE}, 
Abolfazl~Khodadadi\orcidicon{0000-0003-4791-8380},~\IEEEmembership{Student Member,~IEEE}, 
Priyanka~Shinde\orcidicon{0000-0002-4854-976X},~\IEEEmembership{Student Member,~IEEE}, 
Evelin~Blom\orcidicon{0000-0002-8905-3277},~\IEEEmembership{Student Member,~IEEE},
Mohammad~Reza~Hesamzadeh\orcidicon{0000-0002-9998-9773},~\IEEEmembership{Senior Member,~IEEE},
and
Lennart~S\"oder\orcidicon{0000-0002-8189-2420},~\IEEEmembership{Senior Member,~IEEE}
% <-this % stops a space
\vspace{-10 mm}
}  
\markboth{}{Saeed Nordin \MakeLowercase{\textit{et al.}}: Probabilistic Multi-product Trading in Sequential Intraday and Frequency-Regulation Markets}
 \maketitle

\begin{abstract}
With the increasing integration of power plants into the frequency-regulation markets, the importance of optimal trading has grown substantially. This paper conducts an in-depth analysis of their optimal trading behavior in sequential day-ahead, intraday, and frequency-regulation markets. We introduce a probabilistic multi-product optimization model, derived through a series of transformation techniques. Additionally, we present two reformulations that re-frame the problem as a mixed-integer linear programming problem with uncertain parameters. Various aspects of the model are thoroughly examined to observe the optimal multi-product trading behavior of hydro power plant assets, along with numerous case studies. Leveraging historical data from Nordic electricity markets, we construct realistic scenarios for the uncertain parameters. Furthermore, we then proposed an algorithm based on the No-U-Turn sampler to provide probability distribution functions of cleared prices in frequency-regulation and day-ahead markets. These distribution functions offer valuable statistical insights into temporal price risks for informed multi-product optimal-trading decisions.
\end{abstract}
\IEEEpeerreviewmaketitle

\vspace{-5 mm}
\printnomenclature

\section{Introduction}\label{sec:introduction}
\subsection{Background and Motivation}\label{sec:background_motivation}
As Europe progresses towards establishing a low-carbon power system, several notable structural changes are taking place. These changes encompass an upsurge in the utilization of wind power, the gradual decommissioning of thermal power plants, the implementation of new inter-connectors to enhance exchange capacities, and the establishment of new electricity markets. Concurrently, there is a growing imperative for fast-responding reserves to ensure the security and stability of the power system \cite{rintamaki2020strategic}.

In this evolving landscape, hydro power producers possess a distinct advantage over traditional generators due to their inherent storage capacity and rapid adaptability in production. The ability to store water allows them to optimize revenue generation across multiple markets, presenting a significant challenge and opportunity for these producers \cite{khodadadi2022stochastic}. Furthermore, the increasing reliance on renewable energy sources, such as wind and solar power, has heightened the demand for short-term balancing services. Hydro power, particularly when located near intermittent resources, offers a sustainable and viable alternative for providing these essential services. As a result, hydro producers may transition from primarily supplying energy in the day-ahead (DA) market to offering adjustments and balancing services in intraday (ID) and balancing markets, thus emerging as key players in multiple markets. 
To achieve success in this evolving paradigm, they must consider their bidding strategy across all markets as a unified problem, recognizing that commitments made in one market can influence the flexibility and options available in others \cite{aasgaard2019hydropower,mohammadi2022econometric}.
Given their predominant role in the balancing markets, e.g. in Sweden and Norway, hydro power producers tend to assume the role of a price-making asset, significantly impacting market-clearing prices. This behavior is commonly formulated as a bilevel problem in the relevant literature \cite{sinha2017review,rintamaki2020strategic}.

\textit{frequency containment reserve normal (FCR-N)}:
Frequency containment reserve (FCR) markets aim to provide a stable frequency as it is needed to maintain the balance in power system. Although they are activate automatically, they are traded in electricity market to have enough reserve. 
Between the available FCRs, in this paper, frequency containment reserve normal (FCR-N) is studied. It is used to reserve used in the normal operation of power system which is suitable to our study. In countries in which FCR-N market is not available, similar balancing services are available and could be modeled with similar to FCR-N using the same procedure proposed in this paper.
\vspace{-3 mm}
\subsection{Literature Review}\label{sec:literature}
Bilevel modeling has been popularly leveraged in the electricity market literature to model the interaction between two or more entities where one or more of the players involved can be strategic. The players modeled in bilevel problems can be either strategic or non-strategic depending on whether they are on the upper- or lower-level of the model. Several expansion planning models in the long-term electricity market literature follow a bilevel structure to model the interplay between the system operator and generation or consumption companies \cite{tohidi2017sequential,graces2009bilevel,wang2009strategic,wogrin2011generation}.

Most of the bilevel models proposed in the literature have been focused on the day-ahead (DA) electricity market. In \cite{cui2017bilevel}, the energy storage arbitrage revenue is maximized on the upper-level whereas the market-clearing process is modeled at the lower which considers both energy storage and wind power. A bilevel model with revenue and network constraints in the DA market which also includes the effect of inter-temporal constraints associated with generation scheduling, demand-side bidding, and marginal pricing is presented in \cite{7447796}. 
Some of the research works utilizing the bilevel programming structure also consider two different markets mainly DA and balancing markets. In \cite{do2020stochastic}, the participation of an electricity retailer in the DA and real-time markets is modeled as a two-stage process at the upper level, while the distributed renewable energy producers are modeled at the lower level. Several bilevel models are also developed to consider the demand-side perspectives in electricity markets \cite{cui2020optimal,vaya2014optimal,zeng2020bilevel}.

The hydro power planning problems in the literature have been majorly modeled as multistage stochastic programming problems and solved using several decomposition techniques \cite{gjelsvik2010long,catalao2008scheduling,helseth2017assessing,helseth2021assessing}. However, most of these models assume the hydro power to be price takers in the market.
Out of the limited literature on price-making hydro power producers, most of them resort to simplifying assumptions while omitting certain aspects. For example, hydrological balance and topological details of the hydro power have been omitted in \cite{bushnell2003mixed} and \cite{flach2010long}.

A deterministic study on market power in the hydrothermal systems is carried out in \cite{kelman2001market} by considering the residual demand curve (RDC) without taking the transmission constraints into account. Some other works on single hydro power producers include \cite{baslis2011mid} with RDC and \cite{pousinho2012risk} which neglects transmission constraints. \\
A strategic hydro power offering model based on the residual demand curve scenarios is proposed in \cite{zhang2020optimal} where the effect of crossing the forbidden zone is integrated into the model. 
In this paper, we consider multiple strategic and non-strategic hydro power producers and thermal generators participating in the day-ahead, intraday, and FCR-N markets.
\begin{table}[!ht]
\setlength{\aboverulesep}{0pt}
\setlength{\belowrulesep}{0pt}
\centering
\caption{Literature review on various bilevel models in the electricity market literature on hydro power producers}
\begin{tabularx}{\linewidth}{cYYYYYYYYY}
\toprule 
\textsc{Papers} & i & ii & iii & iv & 
v & vi & vii & viii \\ \midrule 
\cite{rintamaki2020strategic} & \checkmark & \checkmark & \xmark & \xmark & 
\checkmark & \checkmark & \xmark & \checkmark \\ 
\rowcolor{Gray}
\cite{bushnell2003mixed}, \cite{kelman2001market}, \cite{loschenbrand2018hydro} & \checkmark& \checkmark& \checkmark& \xmark  & 
\checkmark  & \xmark& \xmark  & \xmark \\ 
\cite{baslis2011mid} & \xmark & \xmark & \checkmark & \checkmark & 
\checkmark & \xmark & \xmark & \xmark \\ 
\rowcolor{Gray}
\cite{pousinho2012risk} & \xmark & \xmark & \checkmark & \checkmark & 
\xmark & \xmark & \xmark & \xmark \\ 
\cite{zhang2020optimal}, \cite{steeger2017dynamic} & \xmark & \xmark & \checkmark & \xmark & 
\checkmark & \xmark & \xmark & \xmark \\ 
\rowcolor{Gray}
\textbf{This paper} & \checkmark & \checkmark & \checkmark & \checkmark &
\checkmark & \checkmark & \checkmark & \checkmark \\
\bottomrule
\end{tabularx}
\begin{tablenotes}
\small
\item i) Model multiple hydro power units, ii) Model thermal producer, iii) Strategic hydro power units, iv) Cascaded hydro power unit, 
v) Day-ahead electricity market, vi) Intraday electricity market, vii) Balancing electricity market, viii) Model transmission constraints.
\end{tablenotes}
\label{tab:literature}
\end{table}

\vspace{-4 mm}
\subsection{Contributions of this paper}
The current paper contributes to the related body of literature as follows:

\textit{C1}:
Strategic operation of hydro power plants in sequentially cleared electricity market setups: day-ahead and FCR-N markets, and the possibility to trade in the intraday market are formulated through stochastic bilevel optimization problem.
Our modeling benefits from bivariate bid curve analysis in which both offer prices and volumes are variables chosen by the strategic producer. The proposed model can be similarly revised for other types of power plants.

\textit{C2}: 
Two Reformulations are proposed to convert our original nonconvex and nonlinear problem into a mixed integer linear programming (MILP) problem which can be efficiently solved using off-the-shelf solvers.
This is handled by McCormick envelop reformulations and replacing the bilinear terms with linear equivalents.

\textit{C3}:
The available historical data from electricity markets are used to generate scenarios for scenario dependent parameters in different years.
Then, we used No-U-Turn sampler based algorithm to calculate probability distribution function (PDF) of the cleared FCR-N and DA market prices.
The PDF of prices are crucial for those who want to operate or invest in trading in ID and FCR-N markets by looking at the optimal price distributions.

\textit{C4}:
A series of case studies are used for concept-proving and testing functionality of our proposed methodology in the sequential market analysis. They study different aspects of the complexities in operation of hydro power plants in the multi-market setups.

Numerical results of the proposed model for the illustrative example above are presented in Section \ref{sec:results:case1} and \ref{sec:results:case2}. Our proposed methodology is explained comprehensively using this illustrative example.
The proposed model is explained in detail for a general problem in the rest of the current section.

\section{Proposed Bilevel Formulation for Hydro-Dominated Power System}\label{sec:bilevel}
\begin{figure}[ht]
\centering
\includegraphics{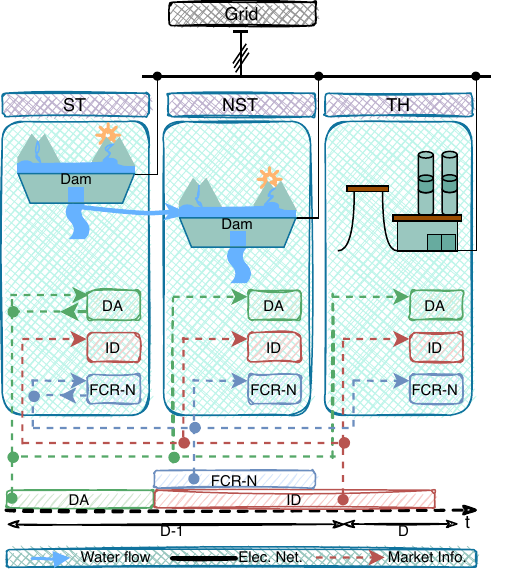}
\caption{Structure of the proposed model. 
% DA: Day-ahead, FCR-N: Frequency containment normal reserve, ID: Intraday, ST: Strategic unit, NST: Non-strategic unit, TH: Non-strategic thermal unit.
}
\label{fig:structure}
\end{figure}

In this section, we have extended the formulation of the illustrative example to propose a bilevel problem to find an optimal solution for the concurrent operation of strategic hydro power producers in the DA, FCR-N, and ID electricity markets. These strategic units are price-makers in the DA and FCR-N markets but they are price-takers in the ID market.
The structure of the proposed model is shown in Fig. \ref{fig:structure}.

\begin{figure}[ht]
\centering
\includegraphics{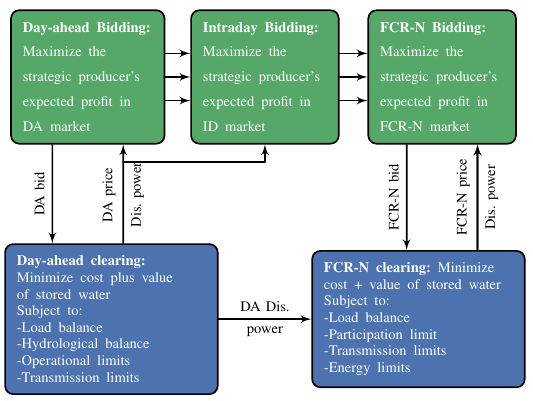}
\vspace{-6 mm}
\caption{Schematic of the proposed bilevel programming problem. 
DA: Day-ahead, ID: Intraday, FCR-N: Frequency containment normal reserve. Dis.: Dispatched.
}
\label{fig:schematic}
\end{figure}

For the sake of clarity, the interaction of units and different markets with their respective constraints are depicted in Fig. \ref{fig:schematic} and elaborated in the following subsections.

\vspace{-4 mm}
\subsection{Upper-level problem}
The upper-level problem formulation is written in \eqref{eq_detail_Obj} to \eqref{eq_detail_feasFCRNclearing} for set of variables $\setXup=\{\varphDA$,$\varphDAseg$, $\varpBIDhDA$, $\varpBIDhFC$, $\varphFC$, $\varphFCseg$,$\varphIDp$, $\varphIDm$, $\varplDA$,
$\varqhDA$, $\varBIDhDAseg$, $\varBIDhFCseg$, $\vardh\}$ in which $\indn\in\setN=\{1$, $2$, $\dots$, $\conNHS\}$ as the node index, 
$\indt\in\setT=\{1$, $2$, $\dots$, $\conNT\}$ as the time index, 
$\indw\in\setW=\{1$, $2$, $\dots$, $\conNW\}$ as the scenario index, 
$\indk\in\setK=\{1$, $2$, $\dots$, $\conNHS\}$ as the hydro generation segment index, 
$\indl\in\setL=\{1$, $2$, $\dots$, $\conNL\}$ as the line index, 
$\inds\in\setS=\{1$, $2$, $\dots$, $\conNBS\}$ as the price and volume segments. 
$\setYDA$ and $\setYFC$ are defined as the set of optimal solutions for the lower-level DA and FCR-N markets, respectively.
The objective function of this problem in \eqref{eq_detail_Obj} is to maximize the total revenue of strategic hydro units in the DA, FCR-N, and ID markets plus the total value of stored water. The objective function is multiplied by its probability $\parprob$ and summed over scenarios as the proposed problem is a stochastic problem. In this formulation, $\setNST\in\{1$, $2$, $\dots$, $\conNST\}$ is the set of strategic units, $\setNNST\in\{1$, $2$, $\dots$, $\conNNST\}$ is the set of non-strategic hydro units, $\setNNSTTH\in\{1$, $2$, $\dots$, $\conNNSTTH\}$ is the set of non-strategic thermal units.

In the Nordic electricity market, once the DA market is cleared, the DA market clearing price and dispatched power are available for ID market participants. Looking at the historical data, we see a strong correlation between the DA market and ID market prices (mostly linear relation). Therefore, this relation between ID market prices ($\parlambdahIDp$ and $\parlambdahIDm$) and the DA market price $\varduallambdahDA$ is determined with statistical analysis. The statistical properties of the DA and ID prices and their linear relation are provided in the numerical results of this paper. Power volumes in ID market are relatively smaller than DA market prices (about 1-7\%). Therefore, the volume limit, $\parGHIDmax$, in the ID market is determined from the dispatched powers in the DA market $\varphDA$ and $\varphDAseg$.
\begin{subequations}
\begin{align}
\SwapAboveDisplaySkip
&\underset{\setXup, \setYDA, \setYFC}{\text{\normalsize Maximize}}\; {\textstyle \sum}_{\indw} \parprob (\underset{\indn\in\setNST,\inds,\indt}{\textstyle \sum} \underbrace{\varphDAseg\varduallambdahDA}_{\expEone}+\nonumber\\
&\underset{\indn\in\setNST,\inds,\indt}{\textstyle \sum}\underbrace{\varphFCseg\varduallambdahFC}_{\expEtwo}+\underset{\indn\in\setNST,\indt}{\textstyle \sum}({\parlambdahIDp\varphIDp-\parlambdahIDm\varphIDm})+\nonumber\\
&\parlambdaf{\textstyle \sum}_{\indn\in\setNST}\varmhT{\textstyle \sum}_{\indj\in\setN}\parAD\parpammah); \label{eq_detail_Obj}
\end{align}
\end{subequations}

The above objective function is subjected to the following constraints. The balance between production and discharge is enforced by \eqref{eq_detail_proddis}.The expression "$\forall\indt,\indw$" is dropped from now on for the sake of brevity.
\begin{equation}
\begin{aligned}
\SwapAboveDisplaySkip
&{\textstyle \sum}_{\indk} \parmuh\vardh = {\textstyle \sum}_{\inds} \varphDAseg + \varphIDp - \varphIDm, \forall \indn \!\in\! \setNST;   \label{eq_detail_proddis}\\
\end{aligned}
\end{equation}

Reservoir content at station $\indn$, spillage from station $\indn$, and discharge volume of station $\indn$ are limited by their minimum or maximum values in \eqref{eq_detail_dischargelim}. 
\begin{equation}
\parMHmin \!\!\le\!\! \varmh \!\!\!\le\!\! \parMHmax\!;\!
\parSHmin \!\!\!\le\!\! \varsh \!\!\!\le\!\! \parSHmax\!;\!
\parQHmin \!\!\le\!\! \vardh \!\!\le\!\! \parQHmax,\!\forall\indn \!\!\in\!\! \setNST \!;\!  
\label{eq_detail_dischargelim}
\end{equation}

Price offer of hydro unit $\indn$ in DA and FCR-N markets are limited by their minimum/maximum bid price in 
\eqref{eq_detail_bidlimitFC}.
\begin{equation}
\begin{aligned}
\SwapAboveDisplaySkip
&\parBIDhDAmin\le\varBIDhDAseg\le\parBIDhDAmax;
\parBIDhFCmin\le\varBIDhFCseg\le\parBIDhFCmax, \forall\indn\in\setNST; \label{eq_detail_bidlimitFC}
\end{aligned}
\end{equation}

The requirements on the FCR-N offers are specified in \eqref{eq_detail_fcrnreq}.
\begin{equation}
    \begin{aligned}
    \SwapAboveDisplaySkip
 & {\textstyle \sum}_{\inds} \varphFCseg \geq (2\parGHmax\pardeltaf)/\pardelta, \forall\indn\in\setNST; \label{eq_detail_fcrnreq}
    \end{aligned}
\end{equation}

To make sure that the bidding curve is descending the constraints
\eqref{eq_detail_bidsegFCminusone} are used.
\begin{equation}
\begin{aligned}
\SwapAboveDisplaySkip
&\varBIDhDAsegminusone\le\varBIDhDAseg;
\varBIDhFCsegminusone\le\varBIDhFCseg, \forall\indn\in\setNST; \label{eq_detail_bidsegFCminusone}
\end{aligned}
\end{equation}

Total power generation at node $\indn$ is limited by maximum/minimum generation capacity in \eqref{eq_detail_maxG2} and \eqref{eq_detail_minG2}.
\begin{subequations}
\begin{align}
\SwapAboveDisplaySkip
& {\textstyle \sum}_{\inds} (\varpBIDhDA+\varpBIDhFC)+\varphIDp-\varphIDm\le\parGHmax, \forall\indn\in\setNST;  \label{eq_detail_maxG2}\\
& {\textstyle \sum}_{\inds} (\varpBIDhDA -\varpBIDhFC)+\varphIDp-\varphIDm\ge 0, \forall\indn\in\setNST; \label{eq_detail_minG2}
\end{align}
\end{subequations}

Buy and sell volumes in the ID market are limited in 
\eqref{eq_detail_bidmaxIDm}, respectively.
\begin{subequations}
\begin{align}
\SwapAboveDisplaySkip
&\!\varphIDp\le\parGHIDmax;
\varphIDm\le\parGHIDmax, \forall\indn\in\setNST; \label{eq_detail_bidmaxIDm}\\
&\!\!\varphDAseg\!,\!\varphDA\!,\!\varplDA\!,\!\vardh\!,\!\varsh\!,\!\varmh \!\!\in\!\! \setYDA;
\varphFCseg,\!\varphFC \!\!\in\!\! \setYFC; \label{eq_detail_feasFCRNclearing}
\end{align}
\end{subequations}

\subsection{DA market clearing}
DA market clearing is formulated in \eqref{eq_setLDA_Obj} to \eqref{eq_setLDA_prodnonstr}. In \eqref{eq_setLDA_Obj}, the objective function is to minimize the cost of procuring the required demand by TSO in the DA market. $\setXDA=\{\varphDAseg $, $\varphDA $, $\varplDA $, $\vardh $, $\varsh $, $\varmh|\varphDAseg\ge0$, $\varphDA\ge0$,
$\vardh\ge0$, $\varsh\ge0$, $\varmh\ge0\}$ is the set of DA market decision variables.
\begin{equation}
\begin{aligned}
\SwapAboveDisplaySkip
&\setYDA:=
\text{\normalsize arg}\underset{\setXDA}{\text{\normalsize min}} 
{\textstyle \sum}_{\inds,\indt,\indn\in\setNST}\parBIDhDAseghat\varphDAseg 
+ \\
&{\textstyle \sum}_{\indt,\indn\in\setNNSTTH} \parctDA\varphDA
-\parlambdaf {\textstyle \sum}_{\indn\in\setNNST} \varmhT {\textstyle \sum}_{\indj} \parAD\parpammah;  \label{eq_setLDA_Obj} \\ 
& \text{Subject to: } \eqref{eq_setLDA_Bal}, \eqref{eq_setLDA_hydBal}, \eqref{eq_setLDA_lim}, \eqref{eq_setLDA_llim}, \eqref{eq_setLDA_gbidlim}, \eqref{eq_setLDA_glim}, \text{and } \eqref{eq_setLDA_prodnonstr};
\end{aligned}
\end{equation}
Power balance in DA market for all stations (strategic and non-strategic) is written in \eqref{eq_setLDA_Bal}.
\begin{equation}
\begin{aligned}
\SwapAboveDisplaySkip
&{\textstyle \sum}_{\inds} \varphDAseg\!+\!\varphDA\!+\! {\textstyle \sum}_{\indl} \parAL\varplDA\!=\!\parDDA\!:\!\varduallambdahDA;  \label{eq_setLDA_Bal}
\end{aligned}
\end{equation}
Hydrological balance constraint is formulated in \eqref{eq_setLDA_hydBal}.
\begin{equation}
\begin{aligned}
\SwapAboveDisplaySkip
&\varmh=\varmhminusone+\parVH-{\textstyle \sum}_{\indk}\vardh-\varsh+\\
& \!\!{\textstyle \sum}_{\indj}\parAU ({\textstyle \sum}_{\indk} \vardhminustau \!\p \varshminustau) \!:\!\vardualetaone, \forall\indn \!\!\in\!\! \setNST\cup\setNNST;   \label{eq_setLDA_hydBal}\\
\end{aligned}
\end{equation}
Discharge volume, reservoir content, and spillage from station $\indn$ are limited in \eqref{eq_setLDA_lim}.
\begin{subequations}\label{eq_setLDA_lim}
\begin{align}
\SwapAboveDisplaySkip
&\parQHmin \!\le\! \vardh \!\le\! \parQHmax:\vardualnutwounder,\vardualnutwoover, \forall\indn\in\setNST\cup\setNNST; \label{eq_setLDA_dlim}\\
&\parMHmin\le\varmh\le\parMHmax:\vardualnuthreeunder,\vardualnuthreeover, \forall\indn\in\setNST\cup\setNNST; \label{eq_setLDA_rlim}\\
&\parSHmin\le\varsh\le\parSHmax:\vardualnufourunder,\vardualnufourover, \forall\indn\in\setNST\cup\setNNST; \label{eq_setLDA_slim}
\end{align}
\end{subequations}
Power of lines $\varplDA$ are limited in \eqref{eq_setLDA_llim}.
\begin{equation}
\begin{aligned}
\SwapAboveDisplaySkip
&-\parglNTCmax\le\varplDA\le\parglNTCmax:\vardualnufiveunder,\vardualnufiveover; \label{eq_setLDA_llim}\\
\end{aligned}
\end{equation}
Dispatched power has to be less than the offered quantity as enforced by \eqref{eq_setLDA_gbidlim}. We have used the hat symbol to show that the upper-level variable is used as a parameter in the lower-level problem.  
\begin{equation}
\begin{aligned}
\SwapAboveDisplaySkip
&0\le\varphDAseg\le\parpBIDhDAhat:\vardualnusixunder,\vardualnusixover, \forall\indn\in\setNST; \label{eq_setLDA_gbidlim}\\
\end{aligned}
\end{equation}
In order to ensure that the dispatched power by non-strategic unit is less than the maximum power generation capacity, we impose \eqref{eq_setLDA_glim}.
\begin{equation}
\begin{aligned}
\SwapAboveDisplaySkip
&0\le\varphDA\le\parGHmax:\vardualnusevenunder,\vardualnusevenover, \forall\indn\in\setNNST\cup\setNNSTTH; \label{eq_setLDA_glim}\\
\end{aligned}
\end{equation}
Using the production equivalent and discharge volume, the dispatched power for the non-strategic unit can be calculated according to \eqref{eq_setLDA_prodnonstr}.
\begin{equation}
\begin{aligned}
\SwapAboveDisplaySkip
&\varphDA-{\textstyle \sum}_{\indk} \vardh\parmuh=0:\vardualetatwo, \forall\indn\in\setNNST; \label{eq_setLDA_prodnonstr}\\
\end{aligned}
\end{equation}
\subsection{KKT of DA market clearing}
The KKT conditions of the DA market clearing is straightforward to obtained and they are not derived here. However, the followings will be used later in the Reformulation 1 and 2 which are stated here: 
\begin{subequations}
\begin{align}
\SwapAboveDisplaySkip
& \parBIDhDAseghat + \varduallambdahDA + \vardualnusixover-\vardualnusixunder = 0:\varphDAseg, \forall\indn\in\setNST;  \label{eq_kktDA_stationary_varphDAsegsetNST}\\
& \vardualnusixover \!(\! \parpBIDhDAhat \m \varphDAseg \!)\! \!=\! 
\vardualnusixunder \!(\! \varphDAseg \!)\! \!=\! 0, \forall\indn\in\setNST;  \label{eq_kktDA_compl_vardualnusixunder}
\end{align}
\end{subequations}

\subsection{FCR-N market clearing}\label{sec:FC}
FCR-N market clearing is formulated in \eqref{eq_setLFC_Obj} to \eqref{eq_setLFC_fcrnreqNNST}. In \eqref{eq_setLFC_Obj}, objective function is to minimize the total cost of procuring the required FCR-N resources over all the units.

There is an important consideration about the terms in the objective function. The value of stored water as the opportunity cost for the non-strategic hydro power plant (HPP) has not been included in the \eqref{eq_setLFC_Obj} while we have it in \eqref{eq_setLDA_Obj}. The main reason is that FCR-N market is a capacity market and it does not include energy activation
or remuneration while in the DA market, the market operator clears the market for the energy activation during the day of operation. Thus, we need to only consider the related operational constraints and variables in the DA market.

However, we need to take into account the opportunity cost of non-strategic HPP; otherwise, it would be evident that TSO activates all of its available capacity first without considering its opportunity cost. Based on the definition of opportunity cost, which is the expected foregone profit of the DA market which is allocated to the capacity market instead of the energy market, our \textit{proposed approach} is to find the expected value of the future electricity price for the day-ahead energy market and set it as the capacity costs for the non-strategic HPP. 
It should be noted that the non-strategic HPP does not exert market power and they only watch the market behavior while strategic HPP exercises the market power to maximize their expected profit. 
Set of decision variables for FCR-N market is $\setXFC=\{$ $\varphFCseg$, $\varphFC|\varphFCseg\ge0$, $\varphFC\ge0\}$.
\begin{equation}
\begin{aligned}
\SwapAboveDisplaySkip
&\setYFC:=\text{\normalsize arg}\underset{\setXFC}{\text{\normalsize min}} \underset{\inds,\indt,\indn\in\setNST}{\textstyle \sum} \parBIDhFCseghat\varphFCseg +\!\!\!\!\underset{\indt,\indn\in\setNNST\cup\setNNSTTH}{\textstyle \sum}\!\!\!\! \parctFC\varphFC; 
\label{eq_setLFC_Obj} \\
& \text{Subject to: } \eqref{eq_setLFC_Bal}, \eqref{eq_setLFC_gbidlim}, \eqref{eq_setLFC_fcrnreqNNST}, \text{and } \eqref{eq_setLFC_prodnonstr};
\end{aligned}
\end{equation}

Power balance in FCR-N market for all stations (strategic and non-strategic) is written in \eqref{eq_setLFC_Bal}.
\begin{equation}
\begin{aligned}
\SwapAboveDisplaySkip
& \!\!  {\textstyle \sum}_{\indn\in\setNST,\inds} \varphFCseg \!\p\! 
{\textstyle \sum}_{\indn\in\setNNSTTH\cup\setNNST}\, \varphFC
 \!\!=\!\! \parDFC\!:\!\varduallambdahFC;  \label{eq_setLFC_Bal}\\
\end{aligned}
\end{equation}

Constraint \eqref{eq_setLFC_gbidlim} is enforced to make sure that the dispatched power for the strategic hydro unit $\indn$ in FCR-N market is less than the offered quantity by the hydro unit $\indn$ in FCR-N market.
\begin{equation}
\begin{aligned}
\SwapAboveDisplaySkip
&\varphFCseg\le\parpBIDhFChat:\vardualthetaone, \forall\indn\in\setNST;  \label{eq_setLFC_gbidlim}\\
\end{aligned}
\end{equation}

The FCR-N requirements for the non-strategic players are included in \eqref{eq_setLFC_fcrnreqNNST}.
\begin{equation}
\begin{aligned}
\SwapAboveDisplaySkip
&\varphFC \geq (2\parGHmax\pardeltaf)/\pardelta:\vardualthetasix, \forall\indn\in\setNNST\cup\setNNSTTH;  \label{eq_setLFC_fcrnreqNNST}
\end{aligned}
\end{equation}

According to \eqref{eq_setLFC_prodnonstr}, the maximum power generation capacity of the non-strategic unit is accounted for while dispatching it in the FCR-N market where the DA dispatch ($\varphDAhat$) is considered to be a parameter.
\begin{subequations}
\begin{align}
\SwapAboveDisplaySkip
&\varphFC+\varphDAhat\leq \parGHmax:\vardualthetaseven, \forall\indn\in\setNNST\cup\setNNSTTH; \label{eq_setLFC_prodnonstr}\\
&\varphFCseg \leq \varphDAseghat:\vardualthetanine, \forall\indn\in\setNST; \label{eq_setLFC_DAFC_varphFCseg}\\
&\varphFC \leq \varphDAhat:\vardualthetaten, \forall\indn\in\setNNST\cup\setNNSTTH; \label{eq_setLFC_DAFC_varphFC}
\end{align}
\end{subequations}

\subsection{KKT conditions for FCR-N market clearing}\label{sec:FC:KKT}
Similarly, the KKT conditions of the FCR-N market clearing is straightforward to obtained and they are not derived here. However, the followings will be used later in the Reformulation 1 and 2 which are stated here: 
\begin{subequations}
\begin{align}
\SwapAboveDisplaySkip
& \varBIDhFCseg + \varduallambdahFC + \vardualthetaone + \vardualthetanine 
= 0:\varphFCseg; \forall\indn\in\setNST \label{eq_kktFC_stationary_varphFCseg}\\
& \vardualthetaone(\parpBIDhFChat - \varphFCseg) = 0, \indn\in\setNST;  \label{eq_kktFC_compl_vardualthetaone}
\end{align}
\end{subequations}

\subsection{Proposed No-U-Turn sampler based algorithm}
\begin{figure}
\centering
\includegraphics[]{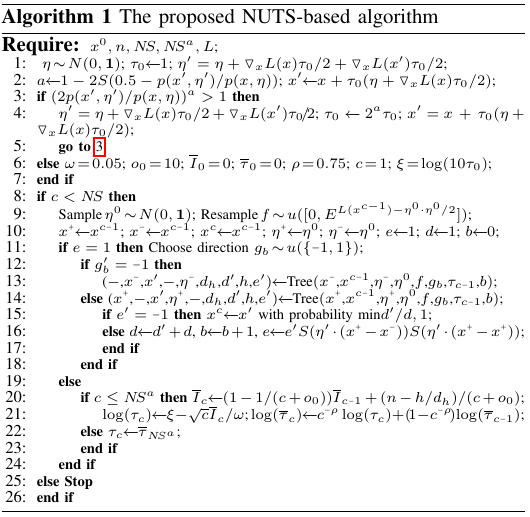}
\makeatletter\def\@currentlabel{1}\makeatother
\label{alg:nuts}
\end{figure}

\begin{figure}
\centering
\includegraphics[]{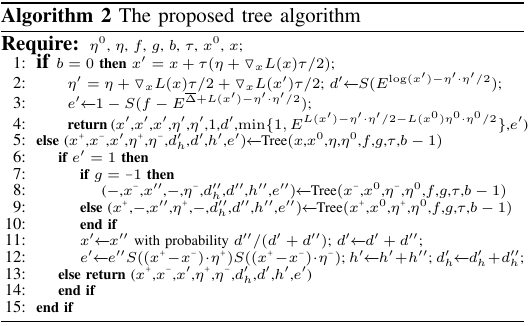}
\makeatletter\def\@currentlabel{2}\makeatother
\label{alg:tree}
\end{figure}

The proposed algorithm based on the No-U-Turn sampler (NUTS) \cite{hoffman2014no} is shown in Algorithm \ref{alg:nuts} and Algorithm \ref{alg:tree}.

\subsection{The Linear Programming (LP) Equivalent}
An approach 
is implemented here to remove the nonlinear terms from \eqref{eq_detail_Obj}. Expressions $\expEone$ and $\expEtwo$ are parts of the stationary conditions. Therefore, they can be replaced by a linear combination of the Lagrangian dual variables. After applying stationary conditions, complementary slackness conditions, Reformulation 1 and Reformulation 2 as they follow, the objective function \eqref{eq_detail_Obj} can be equivalently written as \eqref{eq_detail_Obj_linear}. The original objective function \eqref{eq_detail_Obj} has the bilinear terms $\expEone$ and $\expEtwo$. While they are removed from the linear equivalent \eqref{eq_detail_Obj_linear}. Using \eqref{eq_detail_Obj_linear}, the only remaining non-linear term in the proposed model is $\varphDA\vardualthetaseven$ which will be replaced by $\expEfour$ later.
\begin{subequations}
\begin{align}
\SwapAboveDisplaySkip
&
\substack{
{\text{\normalsize Maximize}}\\
{\setXup, \setYDA, \setYFC}
}
\;
{\textstyle \sum}_{\indw} \parprob (
{\textstyle \sum}_{\indt,\indn\in\setNNSTTH} \parctDA\varphDA
+\nonumber \\
% \displaybreak[0]
&-\!\!\parlambdaf{\textstyle \sum}_{\indn\in\setNNST}\varmhT {\textstyle \sum}_{\indj}\parAD\parpammah \!\!\!+
\!\!{\textstyle \sum}_{\indt,\indn\in\setNNSTTH}\!\parDDA\varduallambdahDA\!+\nonumber \\
% \displaybreak[0]
&\!\!\!\!{\textstyle \sum}_{\indt,\indn\in\setNST\cup\setNNST}(\!\parDDA\varduallambdahDA\!+
\parVH\vardualetaone\!+\!\nonumber \\
% \displaybreak[0]
&{\textstyle \sum}_{\indk}(\parQHmax\vardualnutwoover-
\parQHmin\vardualnutwounder)+
\parMHmax\vardualnuthreeover-
\parMHmin\vardualnuthreeunder+\nonumber \\
% \displaybreak[0]
&\parSHmax\vardualnufourover-
\parSHmin\vardualnufourunder)+
{\textstyle \sum}_{\indt,\indl}(\parglNTCmax\vardualnufiveover+
\parglNTCmax\vardualnufiveunder)+\nonumber \\
% \displaybreak[0]
&\!\!\!\!\underset{\indt,\indn\in\setNNST\cup\setNNSTTH}{\textstyle \sum}\!\!\!\!\!\!\!\!\!\parGHmax\vardualnusevenover
+
\underset{\indt,\indn\in\setNNST\cup\setNNSTTH}{\textstyle \sum}\!\!\!\! \parctFC\varphFC+
\underset{\indt}{\textstyle \sum}\parDFC\varduallambdahFC-\nonumber \\
&\!\!\!\!\!\underset{\indt,\indn\in\setNNST\cup\setNNSTTH}{\textstyle \sum}\!\!\!\!\!\!\!\!\!\vardualthetasix\!(2\parGHmax\pardeltaf)/\pardelta
\!\!+\!\!\!\!\!\!\!\!\!\!\!
\underset{\indt,\indn\in\setNNST\cup\setNNSTTH}{\textstyle \sum}\!\!\!\!\!\!\!\!\!\!\!(\parGHmax\vardualthetaseven-\nonumber \\
&\underbrace{\varphDA\vardualthetaseven}_{\expEfour})+ \underset{\indt,\indn\in\setNNST\cup\setNNSTTH}{\textstyle \sum} \underbrace{\vardualthetaten\varphDA}_{\expEfive}+ \nonumber\\
&\!\!\!\!\!\!\!\underset{\indn\in\setNST,\indt}{\textstyle \sum}\!\!\!\!\!(\parlambdahIDp\varphIDp\!\!-\!\!\parlambdahIDm\varphIDm)\p\parlambdaf\!\!\underset{\indn\in\setNST}{\textstyle \sum}\!\!\varmhT\underset{\indj\in\setN}{\textstyle \sum}\parAD\parpammah); \label{eq_detail_Obj_linear}
\end{align}
\end{subequations}

\textbf{Reformulation 1:} \textit{The nonlinear total revenue of strategic hydro units in DA market ($\expEone$) can be equivalently replaced by a linear sum of upper and lower limits multiplied by their corresponding Lagrangian dual variables.}

\textit{Proof:} First we start with the original bilinear formulation for $\expEone=\varphDAseg\varduallambdahDA$. The aims is to find a linear equivalent for $\expEone$.
First, the strong duality condition for DA market clearing is written in \eqref{eq_kktDA_strong_duality}.
\begin{equation}
\begin{aligned}
\SwapAboveDisplaySkip
& -\!\!\!\!\!\!\!\!\underset{\inds,\indt,\indn\in\setNST}{\textstyle \sum}\!\!\! \parBIDhDAseghat\varphDAseg
\!\!\!-\!\!\!\!\!\!\!\!\underset{\indt,\indn\in\setNNSTTH}{\textstyle \sum}\!\!\! \parctDA\varphDA
\!\!+\!\parlambdaf\!\!\!\!\underset{\indn\in\setNNST}{\textstyle \sum}\!\!\!\varmhT\underset{\indj}{\textstyle \sum}\parAD\parpammah\!\!=\!\! \\ 
&\underset{\indt,\indn\in\setNNSTTH}{\textstyle \sum}\parDDA\varduallambdahDA+\!\!\!\!\!\underset{\indt,\indn\in\setNST\cup\setNNST}{\textstyle \sum}\!\!\!\!\!(\parDDA\varduallambdahDA+
\parVH\vardualetaone+\\
& {\textstyle \sum}_{\indk} (\parQHmax\vardualnutwoover \!\m \parQHmin\vardualnutwounder)+
\parMHmax\vardualnuthreeover \m \parMHmin\vardualnuthreeunder \p \\
&\parSHmax\vardualnufourover \m \parSHmin\vardualnufourunder) \p
 {\textstyle \sum}_{\indt,\indl} (\parglNTCmax\vardualnufiveover \p \parglNTCmax\vardualnufiveunder) \p \\
& {\textstyle \sum}_{\inds,\indt,\indn\in\setNST}  \parpBIDhDAhat\vardualnusixover \p\! {\textstyle \sum}_{\indt,\indn\in\setNNST\cup\setNNSTTH}  \parGHmax\vardualnusevenover;
\label{eq_kktDA_strong_duality}\\
\end{aligned}
\end{equation}

From \eqref{eq_kktDA_stationary_varphDAsegsetNST}, we can write $\varphDAseg\parBIDhDAseghat + \varphDAseg\varduallambdahDA + \varphDAseg\vardualnusixover = 0$;  and from \eqref{eq_kktDA_compl_vardualnusixunder} we have  $\vardualnusixover\parpBIDhDAhat = \vardualnusixover\varphDAseg$.
It gives us $\expEone=\varphDAseg\varduallambdahDA=
-\varphDAseg\parBIDhDAseghat-\vardualnusixover\parpBIDhDAhat$.
Finally the strong duality condition \eqref{eq_kktDA_strong_duality} is rewritten as
${\textstyle \sum}_{\indn\in\setNST,\inds,\indt} {\varphDAseg\varduallambdahDA}=
{\textstyle \sum}_{\indn\in\setNST,\inds,\indt} \expEone=
\m {\textstyle \sum}_{\inds,\indt,\indn\in\setNST} \varphDAseg\parBIDhDAseghat \m {\textstyle \sum}_{\indt,\indn\in\setNST}\!\vardualnusixover\parpBIDhDAhat
=
{\textstyle \sum}_{\indt,\indn\in\setNNSTTH} \parctDA\varphDA
-\parlambdaf{\textstyle \sum}_{\indn\in\setNNST} \varmhT {\textstyle \sum}_{\indj} \parAD\parpammah+
{\textstyle \sum}_{\indt,\indn\in\setNNSTTH} \parDDA\varduallambdahDA+
{\textstyle \sum}_{\indt,\indn\in\setNST\cup\setNNST} (\parDDA\varduallambdahDA+
\parVH\vardualetaone+
{\textstyle \sum}_{\indk} (\parQHmax\vardualnutwoover-
\parQHmin\vardualnutwounder)+
\parMHmax\vardualnuthreeover-
\parMHmin\vardualnuthreeunder+
\parSHmax\vardualnufourover-
\parSHmin\vardualnufourunder)+
{\textstyle \sum}_{\indt,\indl} (\parglNTCmax\vardualnufiveover+
\parglNTCmax\vardualnufiveunder)+
\!\!{\textstyle \sum}_{\indt,\indn\in\setNNST\cup\setNNSTTH}  \parGHmax\vardualnusevenover$;
which is a linear reformulation for $\expEone$.
\hfill$\square$

\textbf{Reformulation 2:} \textit{The nonlinear total revenue of strategic hydro units in FCR-N market ($\expEtwo$) can be equivalently replaced by a linear sum of upper and lower limits multiplied by their corresponding Lagrangian dual variables.}

\textit{Proof:} First we start with the original bilinear formulation for $\expEtwo=\varphFCseg\varduallambdahFC$. The aims is to find a linear equivalent for ${\textstyle \sum}_{\indn\in\setNST,\inds,\indt} {\varphFCseg\varduallambdahFC} = {\textstyle \sum}_{\indn\in\setNST,\inds,\indt} \expEtwo$.

First, the strong duality condition for FCR-N market clearing from Section \ref{sec:FC} is written in \eqref{eq_kktFC_strong_duality}.
\begin{subequations}
\begin{align}
\SwapAboveDisplaySkip
& \m {\textstyle \sum}_{\inds,\indt,\indn\in\setNST} \; \varBIDhFCseg\varphFCseg \m {\textstyle \sum}_{\indt,\indn\in\setNNST\cup\setNNSTTH} \; \parctFC\varphFC= \nonumber \\
&{\textstyle \sum}_{\indt} \parDFC\varduallambdahFC
\!\p\! {\textstyle \sum}_{\inds,\indt,\indn\in\setNST} \; \parpBIDhFChat\vardualthetaone
\!\!\!\!-\!\! \nonumber \\
&\underset{\indt,\indn\in\setNNST\cup\setNNSTTH}{\textstyle \sum}\!\!\!\!\!\!\!\!\!\vardualthetasix(2\parGHmax\pardeltaf)/\pardelta \!\p\!\!\!\!\!\!\!\! \underset{\indt,\indn\in\setNNST\cup\setNNSTTH}{\textstyle \sum} \!\!\!\!\!\!\! (\parGHmax-\varphDAhat)\vardualthetaseven \nonumber \\
& \underset{\indt,\indn\in\setNST}{\textstyle \sum} \vardualthetanine(\varphDAseghat) + \underset{\indt,\indn\in\setNNST\cup\setNNSTTH}{\textstyle \sum} \vardualthetaten(\varphDAhat);
\label{eq_kktFC_strong_duality}
\end{align}
\end{subequations}

Similarly, from \eqref{eq_kktFC_stationary_varphFCseg}, we have $\varphFCseg\varBIDhFCseg + \varphFCseg\varduallambdahFC + \varphFCseg\vardualthetaone + \varphFCseg\vardualthetanine
= 0$; 
and from \eqref{eq_kktFC_compl_vardualthetaone} we have $\vardualthetaone\parpBIDhFChat = \vardualthetaone\varphFCseg$.
It gives us $\expEtwo=\varphFCseg\varduallambdahFC=
-\varphFCseg\varBIDhFCseg - \vardualthetaone\parpBIDhFChat -\varphFCseg\vardualthetanine.
$

Finally the strong duality condition in \eqref{eq_kktFC_strong_duality} is rewritten as
${\textstyle \sum}_{\indn\in\setNST,\inds,\indt}{\varphFCseg\varduallambdahFC}
=
{\textstyle \sum}_{\indn\in\setNST,\inds,\indt}{\expEtwo}
=
-{\textstyle \sum}_{\inds,\indt,\indn\in\setNST} \varphFCseg\varBIDhFCseg -{\textstyle \sum}_{\inds,\indt,\indn\in\setNST} \vardualthetaone\parpBIDhFChat
=
{\textstyle \sum}_{\indt,\indn\in\setNNST\cup\setNNSTTH} \parctFC\varphFC+
{\textstyle \sum}_{\indt} \parDFC\varduallambdahFC-
{\textstyle \sum}_{\indt,\indn\in\setNNST\cup\setNNSTTH} \vardualthetasix(2\parGHmax\pardeltaf)/\pardelta
+
{\textstyle \sum}_{\indt,\indn\in\setNNST\cup\setNNSTTH} (\parGHmax-\varphDAhat)\vardualthetaseven
+{\textstyle \sum}_{\indt,\indn\in\setNNST\cup\setNNSTTH} \,\vardualthetaten(\varphDAhat)$
% +\underset{\indt,\indn\in\setNST}{\textstyle \sum}\parGHmax\vardualthetaeight
which is a linear reformulation for $\expEtwo$.
\hfill$\square$

The McCormic envelopes could be used to have a convex relaxation for $\expEfour={\varphDA\vardualthetaseven}$ where 
$0\le\varphDA\le\parGHmax$ and 
$0\le\vardualthetaseven\le\parmthetaseven$ which are written in \eqref{eq_detail_mccormic_expEfour}. By using $\expEfour$ instead of $\varphDA\vardualthetaseven$ in \eqref{eq_detail_Obj_linear}, the proposed model becomes an MILP problem which can be solved with the commercially available solvers.
\begin{subequations}\label{eq_detail_mccormic_expEfour}
\begin{align}
\SwapAboveDisplaySkip
&\expEfour \ge 0, \indn\in\setNNST\cup\setNNSTTH; \\
&
\expEfour \!\!\ge\! \parGHmax\!\vardualthetaseven \!\p \parmthetaseven\varphDA \!\m\! \parGHmax\parmthetaseven \!\!\!, \indn \!\in\!\! \setNNST \!\cup\! \setNNSTTH; \\
&\expEfour \!\!\le\! \parGHmax\!\vardualthetaseven \;;\!
\expEfour \!\!\le\! \parmthetaseven\!\varphDA \!\!, \indn\in\setNNST\cup\setNNSTTH;
\end{align}
\end{subequations}

The McCormic envelopes could be used to have a convex relaxation for $\expEfive={\varphDA\vardualthetaten}$ where 
$0\le\varphDA\le\parGHmax$ and 
$0\le\vardualthetaten\le\parmthetaten$ which are written in \eqref{eq_detail_mccormic_expEfive}. By using $\expEfive$ instead of $\varphDA\vardualthetaten$ in \eqref{eq_detail_Obj_linear}, the proposed model becomes an MILP problem.
\begin{subequations}\label{eq_detail_mccormic_expEfive}
\begin{align}
\SwapAboveDisplaySkip
&\expEfive \ge 0, \indn\in\setNNST\cup\setNNSTTH; \\
&
\expEfive \!\!\ge\! \parGHmax\!\vardualthetaten \!\p \parmthetaten\varphDA \!\m\! \parGHmax\parmthetaten \!\!\!, \indn \!\in\!\! \setNNST \!\cup\! \setNNSTTH; \\
&\expEfive \!\!\le\! \parGHmax\!\vardualthetaten;
\expEfive \!\!\le\! \parmthetaten\!\varphDA \!\!, \indn\in\setNNST\cup\setNNSTTH;
\end{align}
\end{subequations}

\subsection{Data and model parameters} \label{sec:data}
Scenario dependent parameters are 
$\parVH$, $\varmhzero$, $\parDDA$, $\parctFC$, $\parDFC$, $\parlambdahIDp$, and $\parlambdahIDm$. 
Market data from Nord Pool are used to generate the scenario. 
The hydro data, $\parVH$ and $\varmhzero$, are from Ljungan historical data. The data for stations 1, 2, 3, 4, 5, and 6 is collected from stations Fl{\aa}sj{\"o}n-Grucken, L{\"a}nn{\"a}ssj{\"o}n, R{\"a}tan, Turinge, Bursn{\"a}s, and Havern-Mellansj{\"o}n, respectively.

\vspace{-2 mm}
\section{Numerical Results}\label{sec:results}
Numerical results of the proposed model are discussed in six case studies. Overview of the case studies are listed in Table \ref{tab:results:overview}. Cases I to Case V are used to assess behavior of the proposed model in different conditions and when the size of the problem changes. 
\vspace{-3 mm}
\begin{table}[!ht]
\setlength{\aboverulesep}{0pt}
\setlength{\belowrulesep}{0pt}
\setlength{\tabcolsep}{1.5pt}
\caption{Overview of the model setup in the case studies}\label{tab:results:overview} \centering
\vspace{- 3 mm}
\begin{tabularx}{\linewidth}{lXXXXXXX>{\arraybackslash}m{3.7cm}}
\toprule 
\parbox[t]{1mm}{\rotatebox[origin=c]{90}{\textsc{Case}}}
&$\conNN     $ %Number of nodes
& NST %$\conNNST   $ %Number of non-strategic HPP N\!N^{(\!\m \labST\!)}}
& ST % $\conNST    $ %Number of strategic HPP N\!N^{(\!\labST\!)}}
& TH %$\conNNSTTH $ %Number of non-strategic thermal units N\!N^{(\!\labTH\!)}}
&$\conNT     $ %Number of time steps
&$\conNW     $ %Number of scenarios
&$\conNL     $ %Number of lines
% &$\conNBS    $ %Number of bid segments
% &$\conNHS    $ %Number of hydro segments
& \textsc{Demonstration goal}
\\ 
\midrule 
I
&$ 3 $ %Number of nodes
&$ 1 $ %Number of non-strategic HPP N\!N^{(\!\m \labST\!)}}
&$ 1 $ %Number of strategic HPP N\!N^{(\!\labST\!)}}
&$ 1 $ %Number of non-strategic thermal units N\!N^{(\!\labTH\!)}}
&$ 1 $ %set Number of time steps
&$ 1 $ %Number of scenarios
&$ 2 $ %Number of lines
% &$  $ %Number of bid segments
% &$  $ %Number of hydro segments
& Illustrate Market Clearing
\\
\rowcolor{Gray}
II
&$ 3 $ %Number of nodes
&$ 1 $ %Number of non-strategic HPP N\!N^{(\!\m \labST\!)}}
&$ 1 $ %Number of strategic HPP N\!N^{(\!\labST\!)}}
&$ 1 $ %Number of non-strategic thermal units N\!N^{(\!\labTH\!)}}
&$ 1 $ %set Number of time steps
&$ 1 $ %Number of scenarios
&$ 2 $ %Number of lines
% &$  $ %Number of bid segments
% &$  $ %Number of hydro segments
& Illustrate Transmission Network
\\
III
&$ 3 $ %Number of nodes
&$ 1 $ %Number of non-strategic HPP N\!N^{(\!\m \labST\!)}}
&$ 1 $ %Number of strategic HPP N\!N^{(\!\labST\!)}}
&$ 1 $ %Number of non-strategic thermal units N\!N^{(\!\labTH\!)}}
&$ 3 $ %set Number of time steps
&$ 1 $ %Number of scenarios
&$ 2 $ %Number of lines
% &$ 2 $ %Number of bid segments
% &$ 2 $ %Number of hydro segments
& Market Interaction of HPP and Water Value
\\ 
\rowcolor{Gray}
IV
&$ 3 $ %Number of nodes
&$ 1 $ %Number of non-strategic HPP N\!N^{(\!\m \labST\!)}}
&$ 1 $ %Number of strategic HPP N\!N^{(\!\labST\!)}}
&$ 1 $ %Number of non-strategic thermal units N\!N^{(\!\labTH\!)}}
&$ 24 $ %set Number of time steps
&$ 1 $ %Number of scenarios
&$ 2 $ %Number of lines
% &$ 2 $ %Number of bid segments
% &$ 2 $ %Number of hydro segments
& Market Power Exercise
\\ 
V
&$ 118 $ %Number of nodes
&$ 14 $ %Number of non-strategic HPP N\!N^{(\!\m \labST\!)}}
&$ 1 $ %Number of strategic HPP N\!N^{(\!\labST\!)}}
&$ 4 $ %Number of non-strategic thermal units N\!N^{(\!\labTH\!)}}
&$ 24 $ %set Number of time steps
&$ 20 $ %Number of scenarios
&$ 117 $ %Number of lines
% &$ 4 $ %Number of bid segments
% &$ 4 $ %Number of hydro segments
& Large Scale
\\ 
\bottomrule
\end{tabularx}
\begin{tablenotes}
\small
\item 
NN: Number of buses ($\conNN$);
NST: Number of non-strategic HPP ($\conNNST$); %N\!N^{(\!\m \labST\!)}}
ST: Number of strategic HPP ($\conNST$); %N\!N^{(\!\labST\!)}}
TH: Number of non-strategic thermal units ($\conNNSTTH$);
NT: Number of time steps ($\conNT$);
NW: Number of time scenarios ($\conNT$);
NL: Number of lines ($\conNL$);
\end{tablenotes}
\end{table}

\vspace{-3 mm}
\subsection{Case I: (Illustrate Market Clearing)}\label{sec:results:case1}
A simplified version of the proposed model is used in this section to illustrate DA and FCR-N market clearing. 
All the constraints related to ID market and water flow are removed from detail formulation in Section \ref{sec:bilevel}. As listed in Table \ref{tab:results:overview}, for simplification, we have assumed that all sets of indices except $\setN$ and $\setL$ have one member. Also, there is one scenario with probability one. There are three units connected together. Units 1, 2, and 3 are ST, NST, and TH, respectively.
Generation portfolio and market clearing prices with high, medium, and low demands are presented in Table \ref{tab:results:case1}. There are two main parts with/without FCR-N market.

\textit{DA clearing without FC}: 

\textit{Low demand}:
NST unit is cleared by the market operator due to its zero marginal cost. Therefore, DA prices $\varduallambdahDA$ are all zero. 

\textit{Medium demand}:
In DA market, the demand is higher than the total capacity of NST unit and ST unit takes over the generation but it bids as high as the cost of TH unit to make sure it is cleared. Therefore, the price is 15 \EUR{}/MWh. 

\textit{High demand}:
In DA market, demand is more than the total capacity of ST and NST units which results in the dispatch of TH unit. The ST unit is the price-maker and bids as much as possible, which is the price cap of \parpricecapDA \EUR{}/MWh. 
Therefore, even with a smaller power generation, it can earn more compared to the previous case, $20\!\times\!200 \!>\! 100\!\times\!15$. 

\textit{DA clearing with FC}: 
The DA demands are the same as the previous part but the FCR-N demand is 20MW. 

\textit{Low demand}:
The ST unit bids the required demand in the DA market to have enough margin to be dispatched in the FCR-N market  \eqref{eq_setLFC_prodnonstr}.

\textit{Medium demand:}
In this demand level, after dispatching all the capacity of NST unit, ST unit takes over the generation by bidding up to the TH unit variable cost and reserving some capacity for FCR-N market. Thus, TH unit is the price-maker in the DA market, while in the FCR-N market, ST unit is the marginal producer and sets the prices to the price cap.

\textit{High demand:}
In this demand level, NST and TH units are dispatched in full in the DA market and that makes the ST unit the price-maker for both DA and FCR-N markets.
\begin{table}[!ht]
\setlength{\aboverulesep}{0pt}
\setlength{\belowrulesep}{0pt}
\centering
\setlength{\tabcolsep}{2.5 pt}
\caption{Generation Portfolio and Market Clearing Prices in Case I
}
\vspace{-3mm}
\begin{tabularx}{\linewidth}{c|c|Ycccccc}
\toprule 
& & \multicolumn{3}{c}{\textsc{Demand}} & \multicolumn{2}{c}{\textsc{Generation}} & \multicolumn{2}{c}{$\lambda$} \\
& & & & & & &
\\[-2.3ex]
\cmidrule(lr){3-5} 
\cmidrule(lr){6-7} 
\cmidrule(lr){8-9}
\\[-2.3ex]
& & & DA & FCR-N & DA & FCR-N & DA & FCR-N \\
\parbox[t]{1mm}{\multirow{-3}{*}{\rotatebox[origin=c]{90}{\scriptsize \textsc{Network}}}}
&\parbox[t]{1mm}{\multirow{-3}{*}{\rotatebox[origin=c]{90}{FC}}}
& & \![MWh]\! & \![MW]\! & \![MWh]\! & \![MW]\!& \![\!\text{\EUR{}}/MWh]\!& \![\!\text{\EUR{}}/MW]\!
\\ [1.5ex]
\midrule 
& & H & (50, 50, 70)$^{i}$ & NA & (\textbf{20}, 50, 100) & NA & (\parpricecapDA)$^{ii}$ & NA  
\\
\rowcolor{Gray}
& & M & (50, 50, 40) & NA & (90, 50, \textbf{0}) & NA & (15) & NA  
\\
\parbox[t]{1mm}{\multirow{-3}{*}{\rotatebox[origin=c]{90}{\scriptsize w/o Net.}}}
& 
\parbox[t]{1mm}{\multirow{-3}{*}{\rotatebox[origin=c]{90}{\scriptsize w/o FC}}}
& L & (4, 5, 25) & NA & (0, \textbf{34}, 0) & NA & (0) & NA  
\\
\rowcolor{Gray}
\midrule
& & H & (50, 50, 70) & 20$^{iii}$ & (\textbf{20}, 50, 100) & (\textbf{20}, 0, 0) & (\parpricecapDA) & (\parpricecapFC)  
\\
& & M & (50, 50, 40) & 20 & (85, 50, \textbf{5}) & (\textbf{15}, 0, 5) & (15) & (\parpricecapFC)
\\
\rowcolor{Gray}
\parbox[t]{1mm}{\multirow{-3}{*}{\rotatebox[origin=c]{90}{\scriptsize w/o Net.}}}
& 
\parbox[t]{1mm}{\multirow{-3}{*}{\rotatebox[origin=c]{90}{\scriptsize w/ FC}}}
& L & (4, 5, 25) & 20 & (\textbf{34}, 0, 0) & (\textbf{20}, 0, 0) & (0) & (\parpricecapFC)  
\\
\bottomrule
\end{tabularx}
\begin{tablenotes}
\small
\item i) (ST, NST, TH) different values for each unit; ii) (x)=(x,x,x) same for all units; iii) FCR-N market demand is not unit-specific; 
H: High; M: Medium; L: Low; NA: Not applicable; w/ Net.: with NTC limits; w/o: without NTC limits; Bold: marginal producer.
\end{tablenotes}
\label{tab:results:case1}
\end{table}

\begin{figure}
\centering
\includegraphics{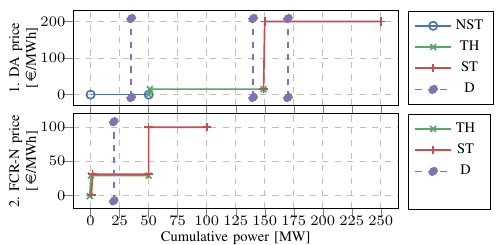}
\caption{Merit order list in Case I (Illustrative Market Clearing).
}
\label{fig:case1:merit}
\end{figure}

Fig. \ref{fig:case1:merit} shows the merit order list of all units in Case I. Capacity of ST, NST, and TH units are 100, 50, and 100 MW, respectively.

\textit{DA price:} The DA market demand (50, 50, 70) MW for (ST, NST, TH) units are scaled while the FCR-N market demand is set to zero.
(1) When the total DA demand is less than 50 MW, NST unit with zero marginal price sets the market price to zero \EUR{}/MWh. It is what happened at L demand in Case I before.
(2) When the total DA demand is higher than 50 MW, ST unit is marginal producer. It leads to price of 15 \EUR{}/MWh for total demands 50 to 150 MW as the ST unit does not bid higher than marginal cost of TH unit $\parctDA=15$ \EUR{}/MWh.
(3) When the total DA demand is higher than 150 MW, NST and TH units can not be the marginal producer as they have reached their maximum capacity. Therefore, ST unit bids to the maximum price of $\parpricecapDA$ \EUR{}/MWh.

\textit{FCR-N price:} The DA market demands for (ST, NST, TH) units are fixed to (44.1, 44.1, 61.8) MW while the FCR-N market demand is changed from zero to 100 MW. Due to \eqref{eq_setLFC_DAFC_varphFCseg} and \eqref{eq_setLFC_DAFC_varphFC}, generation for the DA market should be higher than the FCR-N market.
(1) When the total FCR-N demand is less than 50 MW, NST unit generates maximum amount of $\IparGHmaxindnthree=50$ MW for the DA market and TH unit generates for FCR-N market with marginal price of TH unit $\parctFC=30$ \EUR{}/MWh.
(2) When the total FCR-N demand is between 50 and 100 MW, NST and TH units are not marginal producer as they are in their maximum capacity. Therefore, the ST unit bids to the maximum price of $\parpricecapFC$ \EUR{}/MWh.

\vspace{-3 mm}
\subsection{Case II: (Illustrate Transmission Network)}\label{sec:results:case2}
In this case study, we investigate the effects of the transmission network on results. 
Similarly, generation portfolios and prices are investigated on high, medium, and low demands and results are listed in Table \ref{tab:results:case2}. The net transfer capacity (NTC) of Line 1 is limited to 20 MW while for Line 2 is still 100 MW. 

\textit{DA clearing without FC}:
\textit{Low demand:} Line flows are less than NTC. Which leads to no line congestion and the same results as the case without transmission bottleneck.

\textit{Medium demand:} 
In DA market, the ST unit imports 20MW through Line 1 and TH unit can export it through Line 2.
Hence, the prices for buses 2 and 3 is set to 15\EUR{}/MWh as the variable cost of TH unit. However, the ST unit in bus 1 acts as the price-maker and pushes the price to the price cap \parpricecapDA\EUR{}/MWh due to the congestion in line 1.\\
\textit{High demand:} As the demand grows, the previous situation remains as the TH unit is the marginal producer afterward in Bus 2 and 3 while the unit 1 is the price-maker in the Bus 1. 

\textit{DA clearing with FC}:
\textit{Low demand:} The ST unit bids zero in the DA market and alongside the NST unit, they meet the demand. But as the NST unit dispatch in the FCR-N market is limited to the DA dispatch, the ST unit is the marginal producer and price-maker in the FCR-N market.

\textit{Medium demand:} Activated bottleneck in Line 1 makes the ST unit as the marginal producer in Bus 1 and TH unit in Bus 2 and 3 in the DA market. In FC, the ST unit bids under the variable cost of the TH unit to be dispatched. 

\begin{figure}
\centering
\includegraphics{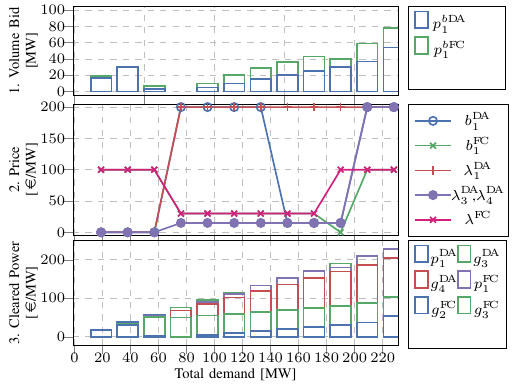}
\caption{Bids of the ST unit in Case II with scaled total demand.}
\label{fig:case2:cumulative_bid_curve}
\end{figure}

\textit{High demand:} In the high demand, the TH unit remains a marginal producer in Bus 2 and Bus 3 in the DA market. In FCR-N market, due to the limited capacity of the TH units, the ST unit acts as the marginal producer and sets the price again to the cap. 
\begin{table}[ht]
\setlength{\aboverulesep}{0pt}
\setlength{\belowrulesep}{0pt}
\centering
\setlength{\tabcolsep}{2pt}
\caption{Generation Portfolio and Market Clearing Prices in Case II 
}
\vspace{-3 mm}
\begin{tabularx}{\linewidth}{c|c|Ycccccc}
\toprule 
& & & \multicolumn{2}{c}{\textsc{Demand}} &  \multicolumn{2}{c}{\textsc{Generation}} & \multicolumn{2}{c}{$\lambda$} \\
 & & & & & & &
\\[-2.3ex]
\cmidrule(lr){3-5}
\cmidrule(lr){6-7} 
\cmidrule(lr){8-9} 
\\[-2.3ex]
& & & DA & FCR-N & DA  & FCR-N & DA & FCR-N \\
\parbox[t]{2 mm}{\multirow{-3}{*}{\rotatebox[origin=c]{90}{\scriptsize  \textsc{Network}}}}
& \parbox[t]{2 mm}{\multirow{-3}{*}{\rotatebox[origin=c]{90}{\scriptsize FC}}}
& & [MWh] & [MW] & [MWh] & [MW] & [\text{\EUR{}}/MWh] & [\!\text{\EUR{}}/MW\!] \\
[1.5ex]
\midrule 
& & H & (50,50,70)$^{i}$ & NA & (\textbf{30}, 50, \underline{90}) & NA & ($\parpricecapDA$,15,15) & NA  \\
\rowcolor{Gray}
& & M & (50,50,40) & NA & (\textbf{30}, 50, \underline{60}) & NA & ($\parpricecapDA$,15,15) & NA  \\
\parbox[t]{1mm}{\multirow{-3}{*}{\rotatebox[origin=c]{90}{\scriptsize w/ Net.}}}
& \parbox[t]{1mm}{\multirow{-3}{*}{\rotatebox[origin=c]{90}{\scriptsize w/o FC}}}
& L & (4,5,25) & NA & (0, \textbf{\underline{34}}, 0) & NA & (0)$^{ii}$& NA  \\ 
\midrule
& & H & (50,50,70) & 20$^{iii}$ & (\textbf{30}, 50, \underline{90}) & (\textbf{\underline{10}}, 0, 10) &  ($\parpricecapDA$,15,15) & ($\parpricecapFC$) \\
\rowcolor{Gray}
& & M & (50,50,40) & 20 & (\textbf{30}, 50, \underline{60}) & (20, 0, \textbf{\underline{0}}) & ($\parpricecapDA$,15,15) & (30)  \\
\parbox[t]{1mm}{\multirow{-3}{*}{\rotatebox[origin=c]{90}{\scriptsize w/ Net.}}}
& \parbox[t]{1mm}{\multirow{-3}{*}{\rotatebox[origin=c]{90}{\scriptsize w/ FC}}}
& L & (4,5,25) & 20 & (24, \textbf{\underline{10}}, 0) & (\textbf{\underline{10}}, 10, 0) & (0) & (\parpricecapFC)  \\
\bottomrule
\end{tabularx}
\begin{tablenotes}
\small
\item i) (ST, NST, TH) different values for each unit; ii) (x)=(x,x,x) same for all units; iii) FCR-N market demand is not unit-specific; H: High; M: Medium; L: Low; NA: Not applicable; w/ Net.: with NTC limits; w/o Net.: without NTC limits; Bold: marginal producer in bus 1; Underline: marginal producer in buses 2 and 3.
\end{tablenotes}
\label{tab:results:case2}
\end{table}
Bid price and volumes of the ST unit in Case II with $\IparglNTCmaxindlone\!=\!20$ MW and $\IparglNTCmaxindltwo\!=\!100$ MW, are shown in FIg. \ref{fig:case2:cumulative_bid_curve}.
Total DA demands (50, 50, and 70 MW for ST, NST, and TH units with generation capacities 100, 50, and 100 MW, respectively) and FCR-N demands of 20 MW are scaled by 0\%, 10\%, \dots, and 120\%.
In total demands less than 57 MW, NST unit is able to generate the DA demand. Therefore, DA price bids are at zero \EUR{}/MWh.
In total demands of 76 MW and above, NST unit generates 50 MW in DA market. It is up to the ST and TH units to generate the remaining. The ST unit bids $\parpricecapDA$ in DA market and $\parctFC$ in FCR-N market and increase the bid volume, accordingly. It leads to different prices at each bus as $\IparglNTCmaxindlone$=20 MW, $\IvarduallambdahDAindnone\!=\!15$ and $\IvarduallambdahDAindntwo\!=\!30$.

\subsection{Case III (Markets Interaction of HPP and Water Value)}\label{sec:results:case3}
\begin{figure}
\centering
\includegraphics{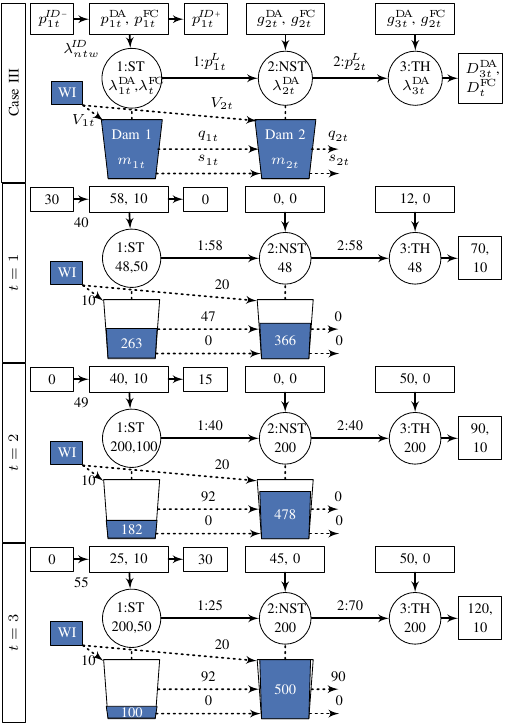}
\label{fig:case3:schematic}
\caption{Schematic and numerical results in Case III. Water flow: dotted arrow; Electric power flow: solid arrow; 
% TO BE ADDED
% ST: strategic HPP; NST: non-strategic HPP; TH: non-strategic thermal power plant; 
WI: water inflow; $\varmhzero=300$ $m^3$.}
\end{figure}

In this case study, we look at interaction of HPPs with the markets considering value of the stored water. Structure and numerical results of the power system and the water network is shown in Fig. \ref{fig:case3:schematic}. 
It includes three HPPs including on ST unit at bus 1, one NST unit at bus 2, and one TH unit at bus 3. Water flows between the reservoirs from bus 1 to bus 2.
We focused on explaining the strategic actions of the target plants. 
Results for congested transmission lines are studied before in Case II.
To simplify, we choose $\parglNTCmax\!=\!200$ MW which is large enough to avoid congestion. Similarly, water flow time from station 1 to downstream station 2, $\partauh$, is set to zero. To see how the impacts of high water value influence the strategic action of ST unit, we have set the water value to a high number for this case.

\textit{Parameters:} 
In this case, different ID price scenarios and load levels have been used to test the reaction of the ST unit in different situations. 
The rest of the parameters are fixed over time as shown in Fig. \ref{fig:case3:schematic}. 
Water inflow to units one and two, $\parVHindnone$ and $\parVHindntwo$, are 10 and 20 $m^3$, respectively; FCR-N demand, $\parDFC$, is 20 MW, thermal costs, $\parctDA$ and $\parctFC$, are 48 and 50 \EUR{}/MWh; future electricity price, $\parlambdaf$, is 26 \EUR{}/MWh and  expected future production equivalent, $\parpammah$ is 0.9 MWh/$m^3$. 

\textit{Results}: As shown in Fig. \ref{fig:case3:schematic}, the ST unit tries to discharge as much as possible to gain more revenue by exercising the market while the market operator seeks to use the water as efficiently as possible and save the water in the NST unit reservoir; however, a proper decision-making framework is required to determine which market is the best option to sell as follows:\\
\textit{Time Step 1:} In time step 1, Strategic Unit 1 submits bids for thermal cost prices in the DA market, competing against Thermal Unit 3 for clearance. To achieve this, Unit 1 procures 30 MW from the ID market due to the favorable relationship between expected ID and DA prices. Consequently, the cleared prices are established at $\IvarduallambdahDA=48$ EUR/MWh and $\IvarduallambdahFC=50$ EUR/MW, mirroring the thermal costs, i.e., $\parctDA=48$ EUR/MWh and $\parctFC=50$ EUR/MWh. Given the high water value in the scenario, the market operator optimizes water resource allocation, leading to the fulfillment of demand by the TH and ST units in the DA market. Additionally, owing to the lower ID price during this period, the ST unit acquires 30 MW to limit discharge and conserve water for future utilization.\\
\textit{Time Step 2:} Subsequently, in time step 2, the projected ID market price surpasses the thermal costs, prompting Strategic Unit 1 to cease procurement from the ID market and instead contribute 15 MW to it. Despite this alteration, the market operator, acknowledging the water's high value, continues to dispatch the TH unit, ensuring water preservation within the NST unit. While this action propels the ST unit to elevate prices to the market cap in both markets, the decision remains advantageous relative to scenarios where the NST unit is dispatched.\\
\textit{Time Step 3:} Time step 3 witnesses a situation where the demand exceeds the combined capacities of the TH and ST units, necessitating the activation of the NST unit. The NST unit's generation capabilities enable its participation in the FCR-N market, causing a transition in the ST unit's role from a price maker to a price taker in this domain. Consequently, the cleared price for the FCR-N market is determined at 50 EUR/MWh, corresponding to the production cost facilitated by the TH unit.

\vspace{-4 mm}
\subsection{Case IV: (Market Power Exercise)}\label{sec:results:case4}
\begin{figure}
\centering
\includegraphics{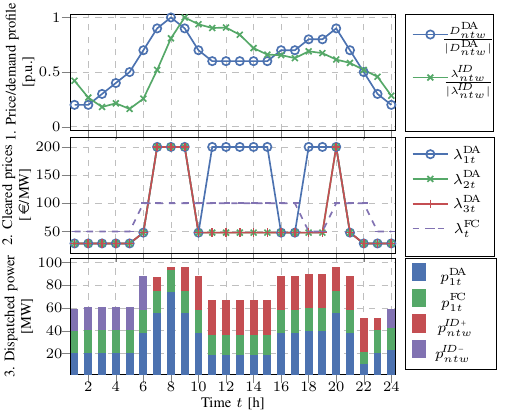}
\vspace{-4mm}
\caption{Numerical results in Case IV (Market Power Exercise).}
\label{fig:case4:profile}
\end{figure}

In this case study, we increase the number of time steps to $\conNT=24$ to assess the performance of participating in the DA, ID, and FCR-N markets, as shown in Fig. \ref{fig:case4:profile}.

Firstly, demand and expected ID price profiles are shown. For the sake of completeness, ID prices fluctuate both in low and high-demand periods.

Secondly, we analyze prices and dispatched power when the water value is relatively low ($\parlambdaf=20$\EUR{}/MWh). Higher demand periods, like time steps 7-9, lead to increased prices in DA markets, primarily due to constrained capacity in NST and TH units. Consequently, the ST unit acts as a price maker during these times. Similar trends are observed in the FCR-N prices, particularly pronounced in bus 1 due to the ST unit and its susceptibility to line congestions.

The ST unit strategically engages with the ID market, purchasing from it when DA prices are low and demand is high (time steps 3 to 5), and selling to it when demand is low and ID prices are high (time steps 11 to 13).
A new variable $\varphDAsegindnone=\sum_{\inds,\indn=1}\varphDAseg$ is used to save space.

% \vspace{-7 mm}
\subsection{Case V: (Large Scale)}\label{sec:results:case5}
\begin{figure}
\centering
\includegraphics{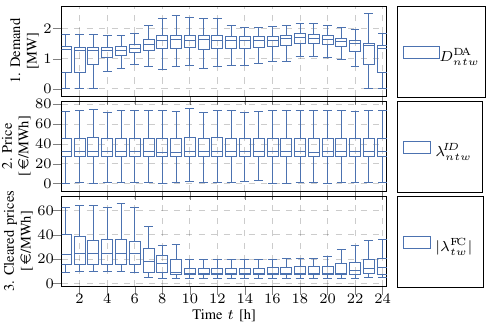}
\vspace{-4mm}
\caption{Boxplot results in Case V (Large Scale).}
\label{fig:case5:boxplot}
\end{figure}

\begin{figure}[!ht]
\centering
\includegraphics[scale=0.55]{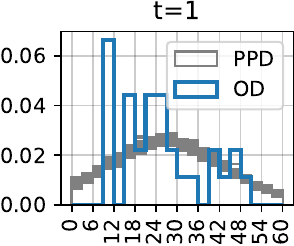}
\includegraphics[scale=0.55]{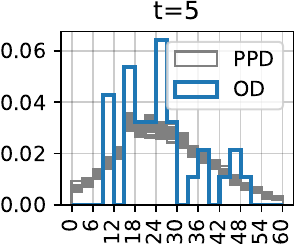}
\includegraphics[scale=0.55]{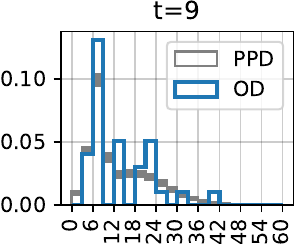}
\includegraphics[scale=0.55]{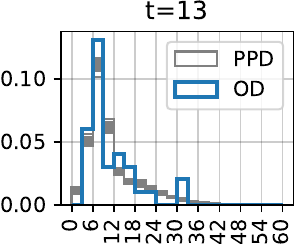}
\includegraphics[scale=0.55]{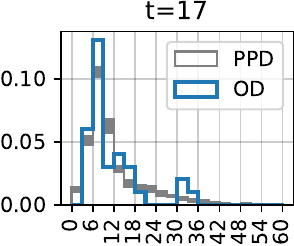}
\includegraphics[scale=0.55]{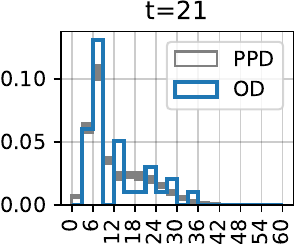}
\vspace{-3 mm}
\caption{PDF of cleared prices of FCR-N market $\varduallambdahFC$ in Case V for different time steps; PPD: Posterior predictive distribution; OD: Observed data.}
    \label{fig:case5:PDF:time}
\end{figure}

The IEEE 118-bus system is used for Case V. It has 186 lines, 4,242 MW loads in 118-buses, and 4,377 MW generators in 18 buses. 
Boxplot results are shown in Fig. \ref{fig:case5:boxplot}.
FCR-N market prices are relatively more volatile at period $\indt=$ 1 to 7 when DA market demands can be relatively lower. 
On the other hand, when there is a higher demand in DA market prices in FCR-N market drop significantly such as $\indt=$ 10 to 19.
The PDFs of cleared prices in the FCR-N market, i.e. $\varduallambdahFC$, in Case V (Large Scale) are shown in Fig. \ref{fig:case5:PDF:time}.

\textit{Firstly}, 
this figure provides probabilistic information about prices for different time steps. 
Such as when there is the highest probability to have a given price range. 
For instance, 
(1) highest probability to have prices 
more than 48 \EUR{}/MWh is at $\indt=1$. 
(2) highest probability to have prices less than 12 \EUR{}/MWh is between 10 and 19.

\textit{Secondly},
looking at the type of PDF functions,
the probability of having prices in range of 30-36 \EUR{}/MWh is similar in $\indt\!=\!1$ and $\indt\!=\!13$. Prices are always less than 36 \EUR{}/MWh at $\indt\!=\!13$ with a narrower PDF. But prices  can vary between 6 and 60 \EUR{}/MWh at $\indt\!=\!1$ with a wider PDF. Therefore, $\indt\!=\!13$ is more promising in terms of more stable FCR-N prices.

\textit{Therefore}, PDF figures help power plants deal with temporal price risks and the decisions are statistically significant and reliable as they are based on the available PDFs using available historical data.

\begin{figure}[hbt!]
\centering
\includegraphics[scale=0.55]{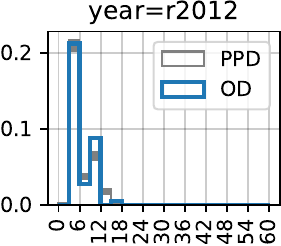}
\includegraphics[scale=0.55]{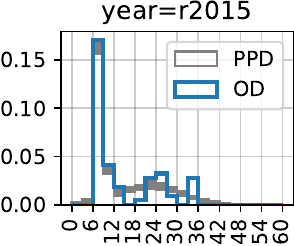}
\includegraphics[scale=0.55]{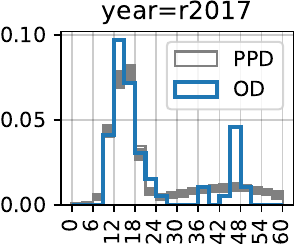}
\includegraphics[scale=0.55]{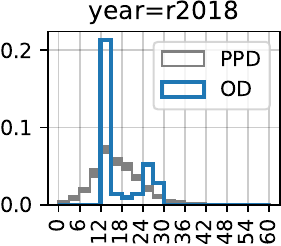}
\includegraphics[scale=0.55]{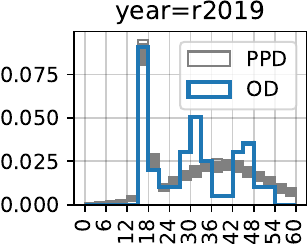}
\includegraphics[scale=0.55]{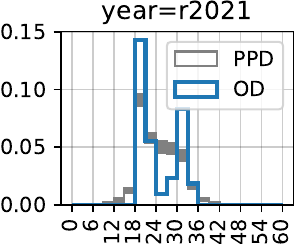}
\vspace{-3 mm}
\caption{PDF of cleared prices of FCR-N market $\varduallambdahFC$ in Case V in different years; PPD: Posterior predictive distribution; OD: Observed data.}
    \label{fig:case5:PDF:year}
\end{figure}

Similarly, PDF of the cleared FCR-N market prices are shown in Fig. \ref{fig:case5:PDF:year}.
There are relatively higher prices in 2017 and 2019 with higher probability in 2019.

\vspace{-7 mm}
\section{Conclusion}\label{sec:conclusion}
\vspace{-1 mm}
We have examined the strategic operation of hydro power plants in sequentially cleared electricity markets: day-ahead, intraday, and frequency-regulation markets.
This helps the power plants optimally trade in multiple markets and manage the available water to generate electricity efficiently.
The power plants and markets are studied under various conditions to investigate market clearing, the market power exercise, and water value.
Available historical market data are used for generating realistic scenarios for uncertain parameters. Subsequently, probability distribution functions (PDFs) of the cleared prices are calculated, which are crucial for power plants to trade in intraday and FCR-N markets.
\vspace{-4 mm}
\ifCLASSOPTIONcaptionsoff
  \newpage
\fi
\bibliographystyle{IEEEtran}
\bibliography{IEEEabrv,bib}

% Generated by IEEEtran.bst, version: 1.14 (2015/08/26)
\begin{thebibliography}{10}
\providecommand{\url}[1]{#1}
\csname url@samestyle\endcsname
\providecommand{\newblock}{\relax}
\providecommand{\bibinfo}[2]{#2}
\providecommand{\BIBentrySTDinterwordspacing}{\spaceskip=0pt\relax}
\providecommand{\BIBentryALTinterwordstretchfactor}{4}
\providecommand{\BIBentryALTinterwordspacing}{\spaceskip=\fontdimen2\font plus
\BIBentryALTinterwordstretchfactor\fontdimen3\font minus
  \fontdimen4\font\relax}
\providecommand{\BIBforeignlanguage}[2]{{%
\expandafter\ifx\csname l@#1\endcsname\relax
\typeout{** WARNING: IEEEtran.bst: No hyphenation pattern has been}%
\typeout{** loaded for the language `#1'. Using the pattern for}%
\typeout{** the default language instead.}%
\else
\language=\csname l@#1\endcsname
\fi
#2}}
\providecommand{\BIBdecl}{\relax}
\BIBdecl

\bibitem{rintamaki2020strategic}
T.~Rintam{\"a}ki, A.~S. Siddiqui, and A.~Salo, ``Strategic offering of a
  flexible producer in day-ahead and intraday power markets,'' \emph{European
  Journal of Operational Research}, vol. 284, no.~3, pp. 1136--1153, 2020.

\bibitem{khodadadi2022stochastic}
A.~Khodadadi, L.~S{\"o}der, and M.~Amelin, ``Stochastic adaptive robust
  approach for day-ahead energy market bidding strategies in hydro dominated
  sequential electricity markets,'' \emph{Sustainable Energy, Grids and
  Networks}, vol.~32, p. 100827, 2022.

\bibitem{aasgaard2019hydropower}
E.~K. Aasg{\aa}rd, S.-E. Fleten, M.~Kaut, K.~Midthun, and G.~A. Perez-Valdes,
  ``Hydropower bidding in a multi-market setting,'' \emph{Energy Systems},
  vol.~10, no.~3, pp. 543--565, 2019.

\bibitem{mohammadi2022econometric}
S.~Mohammadi and M.~R. Hesamzadeh, ``Econometric modeling of intraday
  electricity market price with inadequate historical data,'' in \emph{2022
  IEEE Workshop on Complexity in Engineering (COMPENG)}, July 2022, pp. 1--9.

\bibitem{sinha2017review}
A.~Sinha, P.~Malo, and K.~Deb, ``A review on bilevel optimization: from
  classical to evolutionary approaches and applications,'' \emph{IEEE
  Transactions on Evolutionary Computation}, vol.~22, no.~2, pp. 276--295,
  2017.

\bibitem{tohidi2017sequential}
Y.~Tohidi, M.~R. Hesamzadeh, and F.~Regairaz, ``Sequential coordination of
  transmission expansion planning with strategic generation investments,''
  \emph{IEEE Transactions on Power Systems}, vol.~32, no.~4, pp. 2521--2534,
  2017.

\bibitem{graces2009bilevel}
L.~P. Garces, A.~J. Conejo, R.~Garcia-Bertrand, and R.~Romero, ``A bilevel
  approach to transmission expansion planning within a market environment,''
  \emph{IEEE Transactions on Power Systems}, vol.~24, no.~3, pp. 1513--1522,
  2009.

\bibitem{wang2009strategic}
J.~Wang, M.~Shahidehpour, Z.~Li, and A.~Botterud, ``Strategic generation
  capacity expansion planning with incomplete information,'' \emph{IEEE
  Transactions on Power Systems}, vol.~24, no.~2, pp. 1002--1010, 2009.

\bibitem{wogrin2011generation}
S.~Wogrin, E.~Centeno, and J.~Barquin, ``Generation capacity expansion in
  liberalized electricity markets: A stochastic mpec approach,'' \emph{IEEE
  Transactions on Power Systems}, vol.~26, no.~4, pp. 2526--2532, 2011.

\bibitem{cui2017bilevel}
H.~Cui, F.~Li, X.~Fang, H.~Chen, and H.~Wang, ``Bilevel arbitrage potential
  evaluation for grid-scale energy storage considering wind power and lmp
  smoothing effect,'' \emph{IEEE Transactions on Sustainable Energy}, vol.~9,
  no.~2, pp. 707--718, 2017.

\bibitem{7447796}
R.~Fernández-Blanco, J.~M. Arroyo, and N.~Alguacil,
  ``On the solution of revenue- and network-constrained day-ahead
  market clearing under marginal pricing—part i: An exact bilevel programming
  approach,'' \emph{IEEE Transactions on Power Systems}, vol.~32, no.~1, pp.
  208--219, 2017.

\bibitem{do2020stochastic}
J.~C. Do~Prado and W.~Qiao, ``A stochastic bilevel model for an electricity
  retailer in a liberalized distributed renewable energy market,'' \emph{IEEE
  Transactions on Sustainable Energy}, vol.~11, no.~4, pp. 2803--2812, 2020.

\bibitem{cui2020optimal}
Y.~Cui, Z.~Hu, and H.~Luo, ``Optimal day-ahead charging and frequency reserve
  scheduling of electric vehicles considering the regulation signal
  uncertainty,'' \emph{IEEE Transactions on Industry Applications}, vol.~56,
  no.~5, pp. 5824--5835, 2020.

\bibitem{vaya2014optimal}
M.~G. Vay{\'a} and G.~Andersson, ``Optimal bidding strategy of a plug-in
  electric vehicle aggregator in day-ahead electricity markets under
  uncertainty,'' \emph{IEEE transactions on power systems}, vol.~30, no.~5, pp.
  2375--2385, 2014.

\bibitem{zeng2020bilevel}
B.~Zeng, H.~Dong, F.~Xu, and M.~Zeng, ``Bilevel programming approach for
  optimal planning design of ev charging station,'' \emph{IEEE Transactions on
  Industry Applications}, vol.~56, no.~3, pp. 2314--2323, 2020.

\bibitem{gjelsvik2010long}
A.~Gjelsvik, B.~Mo, and A.~Haugstad, ``Long-and medium-term operations planning
  and stochastic modelling in hydro-dominated power systems based on stochastic
  dual dynamic programming,'' \emph{Handbook of power systems I}, pp. 33--55,
  2010.

\bibitem{catalao2008scheduling}
J.~P. Catalao, S.~J. Mariano, V.~M. Mendes, and L.~A. Ferreira, ``Scheduling of
  head-sensitive cascaded hydro systems: A nonlinear approach,'' \emph{IEEE
  Transactions on Power Systems}, vol.~24, no.~1, pp. 337--346, 2008.

\bibitem{helseth2017assessing}
A.~Helseth, M.~Fodstad, M.~Askeland, B.~Mo, O.~B. Nilsen, J.~I.
  P{\'e}rez-D{\'\i}az, M.~Chazarra, and I.~Guis{\'a}ndez, ``Assessing
  hydropower operational profitability considering energy and reserve
  markets,'' \emph{IET renewable power generation}, vol.~11, no.~13, pp.
  1640--1647, 2017.

\bibitem{helseth2021assessing}
A.~Helseth, M.~Haugen, H.~Farahmand, B.~Mo, S.~Jaehnert, and I.~Stenkl{\o}v,
  ``Assessing the benefits of exchanging spinning reserve capacity within the
  hydro-dominated nordic market,'' \emph{Electric Power Systems Research}, vol.
  199, p. 107393, 2021.

\bibitem{bushnell2003mixed}
J.~Bushnell, ``A mixed complementarity model of hydrothermal electricity
  competition in the western united states,'' \emph{Operations research},
  vol.~51, no.~1, pp. 80--93, 2003.

\bibitem{flach2010long}
B.~C. Flach, L.~Barroso, and M.~Pereira, ``Long-term optimal allocation of
  hydro generation for a price-maker company in a competitive market: latest
  developments and a stochastic dual dynamic programming approach,'' \emph{IET
  generation, transmission \& distribution}, vol.~4, no.~2, pp. 299--314, 2010.

\bibitem{kelman2001market}
R.~Kelman, L.~A.~N. Barroso, and M.~F. Pereira, ``Market power assessment and
  mitigation in hydrothermal systems,'' \emph{IEEE Transactions on power
  systems}, vol.~16, no.~3, pp. 354--359, 2001.

\bibitem{baslis2011mid}
C.~G. Baslis and A.~G. Bakirtzis, ``Mid-term stochastic scheduling of a
  price-maker hydro producer with pumped storage,'' \emph{IEEE Transactions on
  Power Systems}, vol.~26, no.~4, pp. 1856--1865, 2011.

\bibitem{pousinho2012risk}
H.~M. Pousinho, J.~Contreras, A.~G. Bakirtzis, and J.~P. Catal{\~a}o,
  ``Risk-constrained scheduling and offering strategies of a price-maker hydro
  producer under uncertainty,'' \emph{IEEE Transactions on Power Systems},
  vol.~28, no.~2, pp. 1879--1887, 2012.

\bibitem{zhang2020optimal}
C.~Zhang and W.~Yan, ``Optimal offering strategy of a price-maker hydro
  producer considering the effects of crossing the forbidden zones,''
  \emph{IEEE Access}, vol.~8, pp. 10\,098--10\,109, 2020.

\bibitem{loschenbrand2018hydro}
M.~L{\"o}schenbrand, W.~Wei, and F.~Liu, ``Hydro-thermal power market
  equilibrium with price-making hydropower producers,'' \emph{Energy}, vol.
  164, pp. 377--389, 2018.

\bibitem{steeger2017dynamic}
G.~Steeger and S.~Rebennack, ``Dynamic convexification within nested benders
  decomposition using lagrangian relaxation: An application to the strategic
  bidding problem,'' \emph{European Journal of Operational Research}, vol. 257,
  no.~2, pp. 669--686, 2017.

\bibitem{hoffman2014no}
M.~D. Hoffman, A.~Gelman \emph{et~al.}, ``The no-u-turn sampler: adaptively
  setting path lengths in hamiltonian monte carlo.'' \emph{J. Mach. Learn.
  Res.}, 2014.

\end{thebibliography}
\vfill

\clearpage

\end{document}